\documentclass[usegraphicx,usenatbib]{mn2e}
\begin{document}

\voffset -0.3in

%%%\slugcomment{DRAFT --- 2008 Apr08h}

\title[HD~72127AB Interstellar Lines]{Spatial and Temporal Variations in Interstellar Absorption toward HD~72127AB\thanks{Based on observations collected at the European Southern Observatory, Chile, under program 72.C-0682, and on observations made with the NASA/ESA {\it Hubble Space Telescope}, which is operated by the Association of Universities for Research in Astronomy, Inc., under NASA contract NAS5-26555.}}

\author[Welty et al.]{Daniel E. Welty$^1$\thanks{Visiting observer, European Southern Observatory}, Thuso Simon$^2$, and L. M. Hobbs$^3$ \\
$^1$University of Chicago, Astronomy and Astrophysics Center, 5640 S. Ellis Ave., Chicago, IL  60637, USA; welty@oddjob.uchicago.edu \\
$^2$University of New Mexico, Department of Physics and Astronomy, Albuquerque, NM  87131, USA; thuso@unm.edu \\
$^3$University of Chicago, Yerkes Observatory, Williams Bay, WI  53191, USA; hobbs@yerkes.uchicago.edu}

\maketitle
\begin{abstract}
New optical spectra of \mbox{Ca\,{\sc ii}} and \mbox{Na\,{\sc i}} toward HD~72127AB provide additional evidence for both spatial and temporal variations in the complex interstellar absorption along the two sight lines; archival UV spectra yield information on the abundances, depletions, and physical conditions in the gas toward HD~72127A.
Similarities in the strengths of various tracers of interstellar material in the two lines of sight suggest that the total hydrogen column densities ($N$ $\sim$ 2.5$\times$10$^{20}$ cm$^{-2}$) and the depletions and ionization in the main components at low LSR velocities also are similar.
Toward HD~72127A, the main components are relatively cool ($T$ $\la$ 900 K), but with depletions resembling those found in warm, diffuse disc clouds; the generally weaker components at higher velocities have much milder depletions, more like those found in halo clouds.
Several trace neutral species -- \mbox{Ca\,{\sc i}}, \mbox{Cr\,{\sc i}}, and \mbox{Fe\,{\sc i}} -- are much stronger toward HD~72127B, however.
The column density of \mbox{Cr\,{\sc i}}, for example, is about 30 times the value determined toward $\zeta$~Oph (the only previous detection of that species in the ISM).
Dielectronic recombination in warmer gas ($T$ $\ga$ 5000 K) may be largely responsible for the enhanced abundances of those trace neutral species toward HD~72127B.
If the main components toward HD~72127AB are associated with material in the Vela SNR, the differences in abundances and physical conditions occur on scales of about 1100 AU.

\end{abstract}

%%%%%%%%%%%%%%%%%%%%%%%%%%%%%%%%%%%% INTRO %%%%%%%%%%%%%%%%%%%%%%%%%%%%%%%%%%%%

\section{Introduction}
\label{sec-intro}

Small-scale (sub-parsec) spatial structure in the Galactic interstellar medium (ISM) has been revealed via a variety of methods:  multi-epoch \mbox{H\,{\sc i}} observations of pulsars (Frail et al. 1994; Stanimirovi\'{c} et al. 2003), VLBI observations of extended extragalactic radio sources (Dieter, Welch, \& Romney 1976; Brogan et al. 2005), optical spectra of interstellar absorption lines toward binary/multiple star systems and star clusters (Kemp, Bates, \& Lyons 1993; Watson \& Meyer 1996; Points, Lauroesch, \& Meyer 2004), and monitoring of the optical/UV absorption-line profiles toward stars either with significant proper motions or located behind disturbed regions (Hobbs et al. 1991; Rollinde et al. 2003; Welty 2007).
Understanding the nature and origin(s) of that small-scale structure in predominantly neutral gas, however, has been difficult.
Observations of variations in \mbox{Na\,{\sc i}} absorption, for example, seemed to imply local hydrogen densities ($n_{\rm H}$) of thousands per cm$^3$ -- similar to the values inferred more directly from the small-scale variations in \mbox{H\,{\sc i}}, but much higher than would be expected for clouds in thermal pressure equilibrium at typical values $n_{\rm H}T$ $\sim$ 2500 cm$^{-3}$K (Jenkins \& Tripp 2001).
In principle, the observed small-scale structures might be due to geometrical (Heiles 1997), statistical (Deshpande 2000), and/or physical (Lauroesch \& Meyer 2003) properties of the interstellar clouds.
Recent studies combining optical and UV absorption-line data have suggested that at least some of the variations reflect differences in physical conditions (e.g., density, ionization) rather than overall hydrogen column density (Lauroesch et al. 1998; Welty 2007) -- but that would not explain the variations seen directly in $N$(\mbox{H\,{\sc i}}) in other cases.
Summaries of recent observational and theoretical studies of small-scale interstellar structure may be found in Haverkorn \& Goss (2007).

Observations of stars behind the Vela supernova remnant (SNR) provided some of the first (and most striking) indications of small-scale spatial and temporal variability in interstellar absorption lines (Thackeray 1974; Hobbs, Wallerstein, \& Hu 1982; Danks \& Sembach 1995; Cha \& Sembach 2000).
Such variations -- seen toward at least seven stars in that region -- are not entirely unexpected, given the presence of gas at relatively high velocities within the SNR (Wallerstein, Silk, \& Jenkins 1980; Danks \& Sembach 1995; Cha \& Sembach 2000). 
The spatial and temporal differences in the optical lines of \mbox{Na\,{\sc i}} and \mbox{Ca\,{\sc ii}} toward the binary system HD~72127AB have been particularly notable (Thackeray 1974; Hobbs et al. 1982, 1991; Welty, Morton, \& Hobbs 1996; Cha \& Sembach 2000).
That system, consisting of a B2 III primary [$V$ = 5.20; $E(B-V)$ = 0.10] and a B2.5 V secondary [$V$ = 7.09; $E(B-V)$ = 0.10] at ($l$,$b$) = (262.57,$-$3.36), is located behind the edge of a filamentary structure seen in H$\alpha$ emission, less than a degree away from the Vela pulsar.
At a distance of about 480 pc, however, HD~72127AB is well behind the SNR, which is at about 250 pc (Cha, Sembach, \& Danks 1999).
The 4.5 arcsec separation between the two stars corresponds to a transverse distance of about 2200 AU at the stellar distance and about 1100 AU at the distance to the SNR.
Thackeray (1974) first noted differences in the very broad, very strong ($\ga$500 m\AA) interstellar \mbox{Ca\,{\sc ii}} K lines toward the two stars; Hobbs et al. (1981) then reported both corresponding differences in the \mbox{Na\,{\sc i}} D lines and temporal changes in the profile of the \mbox{Ca\,{\sc ii}} line toward HD~72127A.
Detailed fits to a series of \mbox{Ca\,{\sc ii}} and \mbox{Na\,{\sc i}} profiles obtained between 1981 and 1988 toward HD~72127A indicated continuing variations in the column densities and/or velocities of many of the components -- particularly those at low LSR velocities (Hobbs et al. 1991); high \mbox{Ca\,{\sc ii}}/\mbox{Na\,{\sc i}} ratios for the higher velocity components are suggestive of significant grain disruption.
Moderate-resolution UV spectra of the two stars obtained with the {\it Hubble Space Telescope} Goddard High Resolution Spectrograph ({\it HST} GHRS) revealed both similarities and differences in the equivalent widths of the lines from several dominant ions; analysis of the \mbox{C\,{\sc i}} fine-structure excitation implied that some of the gas in both sight lines was characterized by fairly high thermal pressures (Wallerstein et al. 1995a; Wallerstein, Vanture, \& Jenkins 1995b).

In this paper, we discuss new optical spectra of both HD~72127A and HD~72127B, obtained in 2003 with the ESO VLT Ultraviolet and Visual Echelle Spectrograph (UVES), and previously unpublished UV spectra of HD~72127A, obtained in 1992 with the {\it HST} GHRS.
The optical and UV spectra reveal continuing temporal changes in the interstellar absorption toward those two stars and provide new information on the abundances and physical conditions in the gas.
In Section~\ref{sec-data}, we describe the observed spectra and our reduction and analysis techniques.
In Section~\ref{sec-disc}, we discuss the elemental abundances and depletions in the gas, the temporal changes in various features, the differences in absorption between the two sight lines, and the remarkably strong absorption from some trace neutral species seen toward HD~72127B.
In Section~\ref{sec-summ}, we provide a summary of our results.

%%%%%%%%%%%%%%%%%%%%%%%%%%%%%%%%%%%% DATA %%%%%%%%%%%%%%%%%%%%%%%%%%%%%%%%%%%%

\section{Data}
\label{sec-data}

%%%%%%%%%%%%%%%%%%%%%%%%%%%%% OBSERVATIONS %%%%%%%%%%%%%%%%%%%%%%%%%%%%%%%%%%%

\subsection{Observations and data processing}
\label{sec-obs}

Optical spectra of HD~72127A and HD~72127B were obtained on 2003 November 14 using the ESO/VLT UT2 telescope and UVES spectrograph (Dekker et al. 2000), during a run devoted primarily to observations of weak atomic and molecular interstellar lines in the Magellanic Clouds (Welty et al. 2006; Welty \& Crowther, in preparation).
The standard dichroic~1 390/564 setting and a slit width corresponding to 0.7 arcsec were used to obtain nearly complete coverage of the wavelength range 3260--6680 \AA\ (on three CCD detectors: two EEV CCD-44 and one MIT/LL CCID-20) at resolutions of about 4.5 km~s$^{-1}$ in the blue and about 4.9 km~s$^{-1}$ in the red.
This setup includes lines from \mbox{Na\,{\sc i}} (U and D doublets at 3302 and 5889/5895 \AA), \mbox{Ca\,{\sc i}} (4226 \AA), \mbox{Ca\,{\sc ii}} (3933/3968 \AA), and \mbox{Ti\,{\sc ii}} (3383 \AA); the strongest lines from CH (4300 \AA), CH$^+$ (4232 \AA), and CN (3874 \AA); and a number of the diffuse interstellar bands.
Standard routines within IRAF were used to remove bias and to divide sections of the 2-D images containing the spectral order(s) of interest by a normalized flat-field derived from quartz lamp exposures.
The 1-D spectra then were extracted from the flat-fielded image segments via the {\sc apextract} routines, using variance weighting (with the appropriate values for read noise and gain for each detector).
Wavelength calibration was accomplished via Th-Ar lamp exposures, which were obtained at the beginning and end of each night, using the thorium rest wavelengths tabulated by Palmer \& Engelman (1983).
The spectra were then normalized via Legendre polynomial fits to the continuum regions surrounding the interstellar (and stellar) absorption lines.
Equivalent widths for the various interstellar absorption features were measured from those normalized spectra.
The empirical signal-to-noise (S/N) ratios measured in the continuum regions are typically $\sim$ 100--200 per half resolution element -- yielding 3$\sigma$ uncertainties or detection limits for weak, unresolved absorption lines of 1--2 m\AA.
Equivalent widths measured for various interstellar lines toward several other Galactic stars observed during the run show very good agreement with previously reported values (Gredel et al. 1991, 1993; Albert et al. 1993; Welty et al. 1999b).
Observations of the bright, nearby, rapidly rotating star $\psi^2$ Aqr were used to check for weak atmospheric absorption features and instrumental artefacts near the interstellar lines of interest.

UV spectra of HD~72127A were obtained in 1992 April and November with the {\it HST} GHRS, under GO programs 2251 and 3993 (L. M. Hobbs, PI).
The data set includes 13 high-resolution (FWHM = 3.3--3.7 km~s$^{-1}$) ECH-B settings, with central wavelengths from 1745 to 2800 \AA, and ten medium-resolution (FWHM = 13--20 km~s$^{-1}$) G160M settings, with central wavelengths from 1135 to 1664 \AA. 
Multiple, slightly offset sub-exposures were obtained at each setting -- using the WSCAN procedure for the ECH-B spectra and the FP-SPLIT procedure for the G160M spectra -- in order to identify and reduce the effects of detector fixed-pattern noise. 
The individual sub-exposures for each wavelength setting were aligned via fits to the interstellar absorption features, cleaned, and co-added; the summed spectra were then normalized via Legendre polynomial fits to the continuum regions.
The empirical S/N ratios in the normalized spectra range from about 35 to 100 per half resolution element for the ECH-B data and from about 80 to 150 for the lower resolution G160M data.
More detailed discussions of the GHRS data and our adopted reduction and analysis procedures may be found in Welty et al. (1999b).

%%%%%%%%%%%%%%%%%%%%% SPECTRA AND EQUIVALENT WIDTHS %%%%%%%%%%%%%%%%%%%%%%%%%%

\begin{figure}  
\includegraphics[width=84mm]{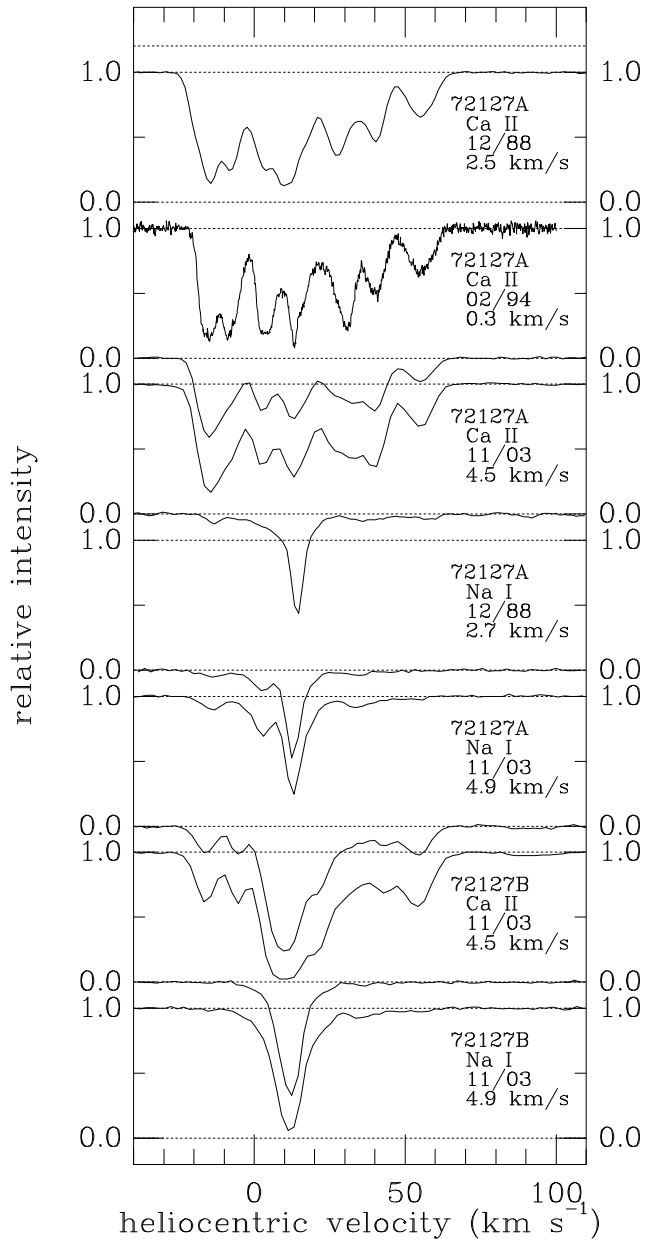}
\caption{Interstellar \mbox{Na\,{\sc i}} and \mbox{Ca\,{\sc ii}} profiles toward HD~72127A and HD~72127B.
For each profile (or pair of profiles), the sight line, species, observation date, and resolution are given at the right.
The UVES spectra (2003 Nov) reveal striking differences between the profiles for the two lines of sight, which are separated by only 4.5 arcsec.
Several earlier profiles toward HD~72127A (Hobbs et al. 1991; Welty et al. 1996) are included to show temporal variations.
Most of the components in \mbox{Ca\,{\sc ii}} exhibit some change over the 15-year span represented here.}
\label{fig:naca}
\end{figure}

\begin{figure}   
\includegraphics[width=84mm]{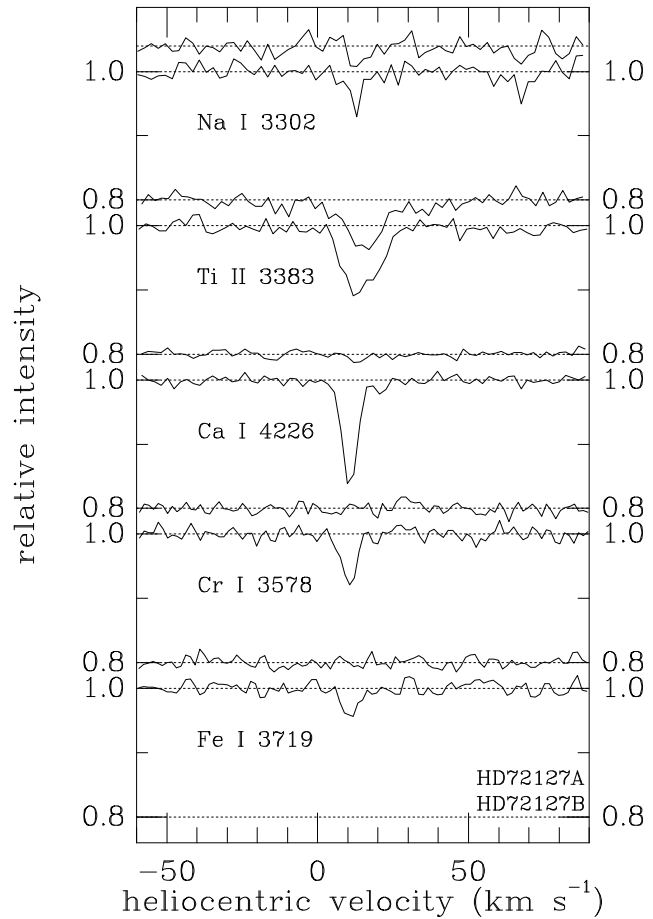}
\caption{Profiles of weaker interstellar lines toward HD~72127A (upper for each pair) and HD~72127B (lower for each pair) observed with UVES.
The weaker member of the \mbox{Na\,{\sc i}} $\lambda$3302 doublet is seen near $+$70 km~s$^{-1}$.
The lines from \mbox{Na\,{\sc i}} and \mbox{Ti\,{\sc ii}} are similar in strength toward the two stars, but the lines from \mbox{Ca\,{\sc i}}, \mbox{Cr\,{\sc i}}, and \mbox{Fe\,{\sc i}} are much stronger toward HD~72127B.}
\label{fig:weak}
\end{figure}

\begin{figure}
\includegraphics[width=84mm]{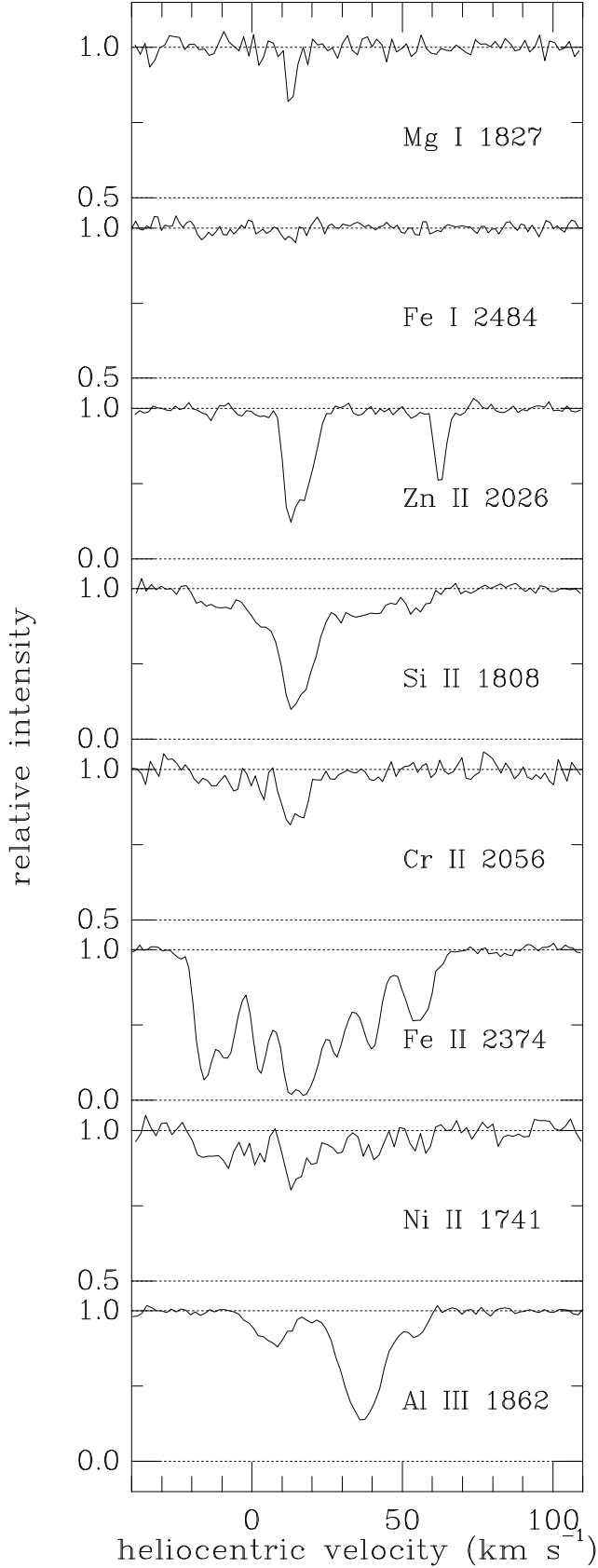}
\caption{Selected UV absorption lines toward HD~72127A, obtained in 1992 with the {\it HST} GHRS, at resolutions of about 3.3--3.7 km~s$^{-1}$.
The absorption near 60 km~s$^{-1}$ in the \mbox{Zn\,{\sc ii}} spectrum is due to \mbox{Mg\,{\sc i}}.
Note the expanded vertical scale for several of the weaker lines.}
\label{fig:uv}
\end{figure}

\begin{table}
\caption{Equivalent widths.}
\label{tab:ewids}
\begin{tabular}{@{}lcrrr}
\hline
\multicolumn{1}{l}{Species}&
\multicolumn{1}{c}{$\lambda$}&
\multicolumn{1}{c}{log($f\lambda$)}&
\multicolumn{1}{c}{HD~72127A}&
\multicolumn{1}{c}{HD~72127B}\\
\hline
\mbox{O\,{\sc i}}   & 1355.5977 &$-$2.803 &$<$2.6         &    --         \\
\mbox{Na\,{\sc i}}  & 3302.3690 &   1.472 &   2.1$\pm$0.9 &   4.0$\pm$1.0 \\
                    & 3302.9780 &   1.170 &   1.9$\pm$0.8 &   2.7$\pm$1.0 \\
                    & 5889.9510 &   3.586 & 216.0$\pm$4.2 & 277.4$\pm$4.2 \\
                    & 5895.9242 &   3.285 & 135.9$\pm$4.0 & 187.6$\pm$3.3 \\
\mbox{Mg\,{\sc i}}  & 1827.9351 &   1.646 &   4.4$\pm$0.8 &    --         \\
                    & 2026.4768 &   2.360 &  18.3$\pm$1.3 &    --         \\
\mbox{Mg\,{\sc ii}} & 1239.9253 &$-$0.106 &  14.8$\pm$1.7 &    --         \\
                    &     --    &    --   & [14.9]        & [13.0]        \\
                    & 1240.3947 &$-$0.355 &   9.5$\pm$1.2 &    --         \\
                    &     --    &    --   &  [8.6]        &  [8.0]        \\
\mbox{Al\,{\sc i}}  & 3944.0060 &   2.664 &$<$1.1         &$<$1.1         \\
\mbox{Al\,{\sc iii}}& 1854.7184 &   3.016 & 169.1$\pm$2.6 &    --         \\
                    & 1862.7910 &   2.714 & 100.6$\pm$2.2 &    --         \\
\mbox{Si\,{\sc i}}  & 2515.0725 &   2.725 &   1.8$\pm$0.4 &    --         \\
\mbox{Si\,{\sc ii}} & 1808.0129 &   0.575 & 115.1$\pm$4.1 &    --         \\
\mbox{S\,{\sc ii}}  & 1250.5780 &   0.832 &[111.5]        &[104.0]        \\
                    & 1253.8050 &   1.136 &[177.1]        &[177.1]        \\
                    & 1259.5180 &   1.320 &[209.0]        &[204.3]        \\
\mbox{Cl\,{\sc i}}  & 1347.2396 &   2.314 &   6.2$\pm$2.2 &    --         \\
\mbox{Ca\,{\sc i}}  & 4226.7280 &   3.874 &   1.0$\pm$0.4 &  16.0$\pm$0.9 \\
\mbox{Ca\,{\sc ii}} & 3933.6614 &   3.392 & 531.8$\pm$1.9 & 508.4$\pm$2.4 \\
                    & 3968.4673 &   3.092 & 327.4$\pm$2.0 & 339.2$\pm$3.6 \\
\mbox{Ti\,{\sc i}}  & 3635.4620 &   2.962 &$<$2.0         &$<$1.6         \\
\mbox{Ti\,{\sc ii}} & 3383.7588 &   3.083 &  17.0$\pm$2.6 &  21.3$\pm$4.2 \\
\mbox{Cr\,{\sc i}}  & 3578.6840 &   3.117 &$<$1.7         &   5.4$\pm$0.7 \\
                    & 3593.4810 &   3.019 &   --          &   5.5$\pm$0.6 \\
                    & 3605.3210 &   2.911 &   --          &   4.5$\pm$0.6 \\
                    & 4254.3320 &   2.670 &   --          &   2.7$\pm$0.4 \\
                    & 4274.7960 &   2.556 &   --          &   1.8$\pm$0.4 \\
                    & 4289.7160 &   2.427 &   --          &   1.2$\pm$0.4 \\
\mbox{Cr\,{\sc ii}} & 2056.2569 &   2.326 &  22.4$\pm$4.5 &    --         \\
                    & 2062.2361 &   2.195 &  17.7$\pm$4.0 &    --         \\
\mbox{Mn\,{\sc i}}  & 4030.7530 &   2.357 &$<$1.2         &$<$1.2         \\
\mbox{Fe\,{\sc i}}  & 2484.0209 &   3.131 &   1.7$\pm$0.5 &    --         \\
                    & 2523.6083 &   2.710 &   0.9$\pm$0.3 &    --         \\
                    & 3440.6057 &   1.910 &   --          &   1.3$\pm$0.6 \\
                    & 3719.9347 &   2.184 &$<$1.3         &   3.5$\pm$0.7 \\
                    & 3859.9114 &   1.923 &   --          &   2.1$\pm$0.4 \\
\mbox{Fe\,{\sc ii}} & 2249.8768 &   0.612 &  35.0$\pm$3.1 &    --         \\
                    & 2374.4612 &   1.871 & 384.0$\pm$3.3 &    --         \\
\mbox{Ni\,{\sc i}}  & 3369.5656 &   1.913 &$<$2.2         &$<$2.6         \\
\mbox{Ni\,{\sc ii}} & 1741.5531 &   1.871 &  32.3$\pm$5.5 &    --         \\
\mbox{Zn\,{\sc ii}} & 2026.1370 &   3.007 &  56.3$\pm$3.6 &    --         \\
                    & 2062.6604 &   2.705 &  42.6$\pm$2.9 &    --         \\
\mbox{Ge\,{\sc ii}} & 1237.0589 &   3.183 &   2.9$\pm$1.1 &    --         \\
                    &     --    &    --   &  [4.0]        &[$<$2.6]       \\
\hline
\end{tabular}
\medskip
~ ~ \\
Entries are equivalent width $\pm$ 1$\sigma$ uncertainty (m\AA); limits are 3$\sigma$.
Wavelengths (in air above 3000 \AA) and $f$-values are from Morton (2003).
Values for lines above 3000 \AA\ are from 2003 UVES spectra; values for lines below 3000 \AA\ are from 1992 GHRS spectra; values in square brackets are from 1992--1993 GHRS spectra (Wallerstein et al. 1995).
\end{table}

\subsection{Spectra and equivalent widths}
\label{sec-ew}

The normalized line profiles for \mbox{Ca\,{\sc ii}} and \mbox{Na\,{\sc i}} toward HD~72127A and HD~72127B are shown in Fig.~\ref{fig:naca}.
In each case, the stronger member of the doublet (if present) is shown at a continuum level of 1.0, and the weaker member of the doublet (if present) is offset by $+$0.2.
Velocities with respect to the local standard of rest (LSR) may be obtained by subtracting 13.4 km~s$^{-1}$ from the heliocentric velocities shown in the figure\footnotemark.
Several previously observed profiles for HD~72127A -- \mbox{Ca\,{\sc ii}} and \mbox{Na\,{\sc i}} from 1988 (FWHM $\sim$ 2.5--2.7 km~s$^{-1}$; Hobbs et al. 1991) and \mbox{Ca\,{\sc ii}} from 1994 (FWHM $\sim$ 0.3 km~s$^{-1}$; Welty et al. 1996) -- are included for comparison. 
Spectra of some of the weaker optical lines are shown in Fig.~\ref{fig:weak}, where the profile for HD~72127A is offset above the one for HD~72127B in each case (and note the expanded vertical scale for all the lines).
Profiles of some of the UV absorption lines toward HD~72127A are shown in Fig.~\ref{fig:uv}, where (again) the vertical scale has been expanded for the weaker lines.
Equivalent widths for the various lines measured from the normalized optical and UV spectra are listed in Table~\ref{tab:ewids}; for comparison, values reported for \mbox{Mg\,{\sc ii}}, \mbox{S\,{\sc ii}}, and \mbox{Ge\,{\sc ii}} by Wallerstein et al. (1995b) (from GHRS spectra obtained in 1992 and 1993) are given in square brackets.
\footnotetext{The conversion to LSR velocities assumes a solar motion of 20 km~s$^{-1}$ toward ($\alpha$,$\delta$) = (18$^{\rm h}$,30\degr).}

In both sight lines, the trace neutral species (e.g., \mbox{Na\,{\sc i}}, \mbox{Mg\,{\sc i}}) are found primarily in an apparently single component at $+$10 to $+$13 km~s$^{-1}$ ($v_{\rm LSR}$ $\sim$ $-$3 to 0 km~s$^{-1}$), with much weaker absorption extending from about $-$20 to $+$55 km~s$^{-1}$.
Toward HD~72127A, the absorption from dominant ions of typically mildly depleted elements (e.g., \mbox{Zn\,{\sc ii}}) is concentrated in several components between about $+$10 and $+$20 km~s$^{-1}$, with hints of much weaker absorption at lower and higher velocities.
The absorption from dominant ions of typically more severely depleted elements (e.g., \mbox{Si\,{\sc ii}}, \mbox{Fe\,{\sc ii}}) also is strongest between $+$10 and $+$20 km~s$^{-1}$, but becomes increasingly prominent at both lower and higher velocities for the species most affected by depletion. 
\mbox{Ca\,{\sc ii}} -- which can be affected by both ionization and depletion (e.g., Welty et al. 1996) -- shows moderately strong absorption over the entire velocity range.

%%%%%%%%%%%%%%%%%%%%%%%%%%%%%% PROFILE FITS %%%%%%%%%%%%%%%%%%%%%%%%%%%%%%%%%%

\subsection{Profile fits}
\label{sec-fits}

Multicomponent fits to observed complex absorption-line profiles enable blended lines and components to be disentangled, account explicitly for saturation effects in the stronger components, and allow spectra obtained at (somewhat) different resolutions to be compared.
The method of profile fitting was therefore used to estimate column densities ($N$), line widths ($b$ $\sim$ FWHM/1.665), and velocities ($v$) for the individual components discernible in the spectra (e.g., Hobbs et al. 1991; Welty, Hobbs, \& Morton 2003).  
For each species, the minimum number of components needed to achieve `satisfactory' fits to the detected lines was adopted (given the resolution and S/N ratios characterizing the spectra and assuming that each component may be represented by a symmetric Voigt profile).  
Multiple lines for a given species (e.g., the two lines of \mbox{Ca\,{\sc ii}}) were fitted simultaneously.
The structure determined for \mbox{Ca\,{\sc ii}} was used to estimate individual component column densities for \mbox{Ti\,{\sc ii}} from the weak $\lambda$3383 lines.
The lines from \mbox{Si\,{\sc ii}}, \mbox{Cr\,{\sc ii}}, \mbox{Fe\,{\sc ii}}, \mbox{Ni\,{\sc ii}}, and \mbox{Zn\,{\sc ii}} seen in the GHRS ECH-B spectra were fitted simultaneously -- using a common velocity structure but adjusting the component column densities (and some $b$-values) for each species.

The adopted interstellar component parameters derived from the UVES spectra of HD~72127A and HD~72127B are listed in Table~\ref{tab:comps}; the component parameters obtained from the simultaneous fit to the GHRS ECH-B spectra of HD~72127A are given in Table~\ref{tab:compuv}.
For each component in each sight line, Table~\ref{tab:comps} gives the component number, the heliocentric velocity and $b$-value (for \mbox{Na\,{\sc i}} and \mbox{Ca\,{\sc ii}}), and the column densities of \mbox{Na\,{\sc i}}, \mbox{Ca\,{\sc i}}, \mbox{Ca\,{\sc ii}}, and \mbox{Ti\,{\sc ii}}.  
Component column densities or $b$-values in square brackets were fixed, but relatively well determined in the fits; values in parentheses also were fixed, but are less well determined.  
For \mbox{Na\,{\sc i}}, the column density of the strongest component in each sight line was fixed at a value consistent with that derived from the weak $\lambda$3302 doublet; the parameters for the other components were derived from fits to the much stronger \mbox{Na\,{\sc i}} D lines.
Independent fits to the profiles of the \mbox{Na\,{\sc i}} D lines and the \mbox{Ca\,{\sc ii}} H and K lines yielded rather similar component velocities (in both sight lines), and components seen in the two species are considered to be associated if the velocities agree within about 1--2 km~s$^{-1}$.

%%%%%%%%%%%%%%%%%%%%%%%%%%%%%%%%%%%%%%% version 2 of comp structure table

\begin{table*}
\begin{minipage}{180mm}
\caption{Component structures (optical data).}
\label{tab:comps}
\begin{tabular}{@{}rrcrrrcrrrr}
\hline
\multicolumn{1}{c}{ }&
\multicolumn{3}{c}{\mbox{Na\,{\sc i}}}&
\multicolumn{1}{c}{\mbox{Ca\,{\sc i}}}&
\multicolumn{3}{c}{\mbox{Ca\,{\sc ii}}}&
\multicolumn{1}{c}{\mbox{Ti\,{\sc ii}}}&
\multicolumn{1}{c}{\mbox{Ca\,{\sc ii}}/\mbox{Na\,{\sc i}}}&
\multicolumn{1}{c}{\mbox{Ca\,{\sc ii}}/\mbox{Ti\,{\sc ii}}}\\
\multicolumn{1}{c}{Comp}&
\multicolumn{1}{c}{$v$}&
\multicolumn{1}{c}{$b$}&
\multicolumn{1}{c}{$N_{10}$}&
\multicolumn{1}{c}{$N_{9}$}&
\multicolumn{1}{c}{$v$}&
\multicolumn{1}{c}{$b$}&
\multicolumn{1}{c}{$N_{10}$}&
\multicolumn{1}{c}{$N_{10}$}\\
\multicolumn{1}{c}{ }&
\multicolumn{1}{c}{(km~s$^{-1}$)}&
\multicolumn{1}{c}{(km~s$^{-1}$)}&
\multicolumn{1}{c}{(cm$^{-2}$)}&
\multicolumn{1}{c}{(cm$^{-2}$)}&
\multicolumn{1}{c}{(km~s$^{-1}$)}&
\multicolumn{1}{c}{(km~s$^{-1}$)}&
\multicolumn{1}{c}{(cm$^{-2}$)}&
\multicolumn{1}{c}{(cm$^{-2}$)}\\
\hline
\multicolumn{11}{c}{HD~72127A} \\
\hline
%%%     v(Na I)       b(Na I)         N(Na I)        N(Ca I)        N(Cr I)       N(Fe I)
%%%     v(Ca II)      b(Ca II)        N(Ca II)       N(Ti II)      Ca II/Na I     Ca II/Ti II
1 &                &             &               &              &  %%           &             &
   $-$27.7$\pm$2.1 &(3.0)        &   1.0$\pm$0.5 &              &               &             \\
%%%
2 &$-$13.8$\pm$0.3 & 4.4$\pm$0.5 &   9.8$\pm$0.6 &              &  %%           &             &
   $-$15.4$\pm$0.1 & 3.2$\pm$0.1 & 218.3$\pm$3.8 &  2.6$\pm$1.4 &  22.3         &$\sim$84.0   \\
%%%
3 &                &             &               &              &  %%           &             &
    $-$8.2$\pm$0.1 &(4.0)        & 112.6$\pm$3.0 &  2.0$\pm$1.5 &               &$\sim$56.3   \\
%%%
4 & $-$1.1$\pm$0.6 &(3.0)        &   8.4$\pm$1.2 &              &  %%           &             &
                   &             &               &              &               &             \\
%%%
5 &    3.7$\pm$0.2 &(2.0)        &  21.3$\pm$1.4 &              &  %%           &             &
       2.9$\pm$0.1 & 3.3$\pm$0.2 & 120.1$\pm$3.0 &  4.7$\pm$1.3 &   5.6         &      25.6   \\
%%%
6 &   12.8$\pm$0.1 & 1.2$\pm$0.1 &[250.0$\pm$70.]&  3.8$\pm$0.7 &  %%           &             &
      12.6$\pm$0.2 & 3.6$\pm$0.4 & 148.4$\pm$12. & 15.9$\pm$1.6 &   0.6         &       9.3   \\
%%%
7 &   18.8$\pm$0.2 &(3.0)        &  14.0$\pm$0.8 &              &  %%           &             &
      18.3$\pm$0.5 &(3.0)        &  41.3$\pm$9.4 & 13.6$\pm$1.5 &   3.0         &       3.0   \\
%%%
8 &                &             &               &              &  %%           &             &
      26.0$\pm$0.2 &(3.0)        &  67.1$\pm$1.8 &  2.8$\pm$1.4 &               &$\sim$24.0   \\
%%%
9 &   32.8$\pm$0.5 &(4.0)        &   7.4$\pm$0.7 &              &  %%           &             &
      32.3$\pm$0.2 &(3.0)        &  81.6$\pm$2.2 &  3.3$\pm$1.4 &  11.0         &$\sim$24.7   \\
%%%
10&   40.7$\pm$1.3 &(4.0)        &   3.0$\pm$0.6 &              &  %%           &             &
      40.0$\pm$0.1 & 3.8$\pm$0.1 & 130.2$\pm$2.9 &  3.0$\pm$1.4 &  43.4         &$\sim$43.4   \\
%%%
11&   52.0$\pm$1.3 &(5.0)        &   2.5$\pm$0.5 &              &  %%           &             &
      54.7$\pm$0.1 & 4.9$\pm$0.1 &  61.6$\pm$1.0 &              &  24.6         &             \\
\hline
\multicolumn{11}{c}{HD~72127B} \\
\hline
%%%     v(Na I)       b(Na I)         N(Na I)        N(Ca I)        N(Cr I)       N(Fe I)
%%%     v(Ca II)      b(Ca II)        N(Ca II)       N(Ti II)      Ca II/Na I     Ca II/Ti II
1 &$-$15.1$\pm$1.3 &(2.0)        &   1.0$\pm$0.4 &              &  %%           &             &
   $-$16.0$\pm$0.1 & 3.5$\pm$0.1 &  58.4$\pm$1.1 &              &  58.4         &             \\
%%%
2 &                &             &               &              &  %%           &             &
    $-$5.4$\pm$0.1 & 2.5$\pm$0.2 &  51.7$\pm$1.2 &  1.7$\pm$1.2 &               &$\sim$30.4   \\
%%%
3 & $-$1.1$\pm$0.4 &(2.5)        &   6.9$\pm$0.7 &              &  %%           &             &
                   &             &               &              &               &             \\
%%%
4 &    6.5$\pm$0.4 &(3.0)        &  31.7$\pm$4.3 &              &  %%           &             &
       4.3$\pm$0.3 &(3.0)        & 113.1$\pm$17. &              &   3.6         &             \\
%%%
5 &   12.2$\pm$0.1 & 2.5$\pm$0.1 &[400.0$\pm$55.]& 57.9$\pm$1.2 & %%1.8$\pm$0.1 & 7.8$\pm$0.6 &
      10.5$\pm$0.1 &(4.0)        &1060.0$\pm$65. & 28.4$\pm$1.9 &   2.7         &      37.3   \\
%%%
6 &   21.4$\pm$0.2 &(3.5)        &  18.2$\pm$0.7 &  5.5$\pm$0.9 &  %%           &             &
      21.0$\pm$0.1 &(4.0)        & 185.7$\pm$3.0 & 20.2$\pm$1.5 &  10.2         &       9.2   \\
%%%
7 &                &             &               &              &  %%           &             &
      29.4$\pm$0.2 &(3.0)        &  41.2$\pm$1.5 &              &               &             \\
%%%
8 &   34.7$\pm$0.4 &(3.0)        &   5.7$\pm$0.5 &              &  %%           &             &
      35.7$\pm$0.3 &(3.0)        &  27.2$\pm$1.3 &              &   4.8         &             \\
%%%
9 &   43.3$\pm$1.2 &(4.0)        &   2.7$\pm$0.5 &              &  %%           &             &
      42.9$\pm$0.2 &(3.0)        &  36.9$\pm$1.2 &              &  13.7         &             \\
%%%
10&   54.2$\pm$1.2 &(4.0)        &   2.1$\pm$0.5 &              &  %%           &             &
      53.8$\pm$0.1 & 5.6$\pm$0.1 &  90.4$\pm$1.6 &  3.5$\pm$1.5 &  43.0         &      25.8   \\
%%%
11&                &             &               &              &  %%           &             &
      88.0$\pm$1.0 &(4.0)        &   4.0$\pm$0.8 &              &               &             \\
%%%
12&                &             &               &              &  %%           &             &
      96.9$\pm$1.3 &(4.0)        &   3.3$\pm$0.8 &              &               &             \\
\hline
\end{tabular}
\medskip
~ ~ \\
The subscript on each column density header indicates the power of ten by which the listed value of that species should be multiplied.
Toward HD~72127B, \mbox{Cr\,{\sc i}} and \mbox{Fe\,{\sc i}} are detected, with [$N$, $v$] equal to [(1.8$\pm$0.1)$\times$10$^{11}$, 11.0$\pm$0.1] and [(7.8$\pm$0.6)$\times$10$^{11}$, 10.6$\pm$0.2], respectively.
\end{minipage}
\end{table*}

\begin{table*}
\begin{minipage}{180mm}
\caption{Component structures (HD~72127A UV data).}
\label{tab:compuv}
\begin{tabular}{@{}rrcrrrrrr}
\hline
\multicolumn{1}{c}{Comp}&
\multicolumn{1}{c}{$v$}&
\multicolumn{1}{c}{$b$}&
\multicolumn{1}{c}{$N$(\mbox{Zn\,{\sc ii}})}&
\multicolumn{1}{c}{$N$(\mbox{Si\,{\sc ii}})}&
\multicolumn{1}{c}{$N$(\mbox{Cr\,{\sc ii}})}&
\multicolumn{1}{c}{$N$(\mbox{Fe\,{\sc ii}})}&
\multicolumn{1}{c}{$N$(\mbox{Ni\,{\sc ii}})}&
\multicolumn{1}{r}{\mbox{Fe\,{\sc ii}}/\mbox{Zn\,{\sc ii}}}\\
\multicolumn{1}{c}{ }&
\multicolumn{1}{c}{(km~s$^{-1}$)}&
\multicolumn{1}{c}{(km~s$^{-1}$)}&
\multicolumn{1}{c}{(10$^{10}$ cm$^{-2}$)}&
\multicolumn{1}{c}{(10$^{13}$ cm$^{-2}$)}&
\multicolumn{1}{c}{(10$^{11}$ cm$^{-2}$)}&
\multicolumn{1}{c}{(10$^{12}$ cm$^{-2}$)}&
\multicolumn{1}{c}{(10$^{12}$ cm$^{-2}$)}&
\multicolumn{1}{c}{ }\\
%%%     v(Fe II)+2.9     b       N(Zn II)        N(Si II)       N(Cr II)        N(Fe II)      N(Ni II)     FeII/ZnII  FeII/NiII  SiII/FeII 
\hline
\multicolumn{1}{r}{$N$(3$\sigma$)} & \multicolumn{2}{c}{ } & 
\multicolumn{1}{c}{9.0} & \multicolumn{1}{c}{3.5} & \multicolumn{1}{c}{4.5} & \multicolumn{1}{c}{2.5} & \multicolumn{1}{c}{2.0} \\
\hline
 1 & $-$24.3$\pm$1.2 & (3.0) &        --     &        --     &              &   1.1$\pm$0.3 &             &    --  \\  %%
 2 & $-$15.9$\pm$0.1 &  2.5  &        --     &   6.6$\pm$0.9 &  6.0$\pm$1.3 &  59.8$\pm$2.8 & 2.9$\pm$0.6 & $>$665 \\  %% 20.6  1.1
 3 &  $-$8.5$\pm$0.2 &  4.0  &        --     &  11.8$\pm$1.1 &  6.8$\pm$1.5 &  52.7$\pm$2.4 & 4.8$\pm$0.7 & $>$585 \\  %% 11.0  2.2
 4 &     2.9$\pm$0.1 &  2.8  &  17.3$\pm$3.8 &  24.6$\pm$1.3 &  5.8$\pm$1.3 &  54.6$\pm$1.4 & 3.4$\pm$0.6 &    315 \\  %% 16.1  4.5
%%%
 5 &    13.2$\pm$0.2 & [3.7] & 338.2$\pm$13. & 111.7$\pm$5.1 & 26.7$\pm$1.9 & 115.1$\pm$7.7 & 6.5$\pm$0.7 &     35 \\  %% 17.7  9.7
 6 &    19.5$\pm$0.3 & [4.0] & 150.5$\pm$7.7 &  58.2$\pm$3.5 &  9.2$\pm$2.1 &  85.1$\pm$1.7 & 3.9$\pm$0.7 &     55 \\  %% 21.8  6.8
%%%
 7 &    28.0$\pm$0.1 &  2.4  &        --     &   7.7$\pm$1.1 &  3.9$\pm$1.4 &  31.1$\pm$1.7 & 2.2$\pm$0.6 & $>$345 \\  %% 14.1  2.5
 8 &    32.9$\pm$0.5 & (3.5) &        --     &  13.7$\pm$1.3 &       --     &  15.1$\pm$1.7 &      --     & $>$170 \\  %%       9.1
 9 &    39.9$\pm$0.1 &  3.4  &   5.0$\pm$3.9 &  12.0$\pm$1.8 &  4.4$\pm$1.4 &  37.4$\pm$1.0 & 3.5$\pm$0.6 &    750 \\  %% 10.7  3.2
 10&    44.6$\pm$1.2 & (3.5) &        --     &   5.5$\pm$1.4 &       --     &        --     &      --     &    --  \\  %%
 11&    54.7$\pm$0.2 &  5.7  &        --     &  16.6$\pm$2.0 &       --     &  32.9$\pm$1.2 & 2.5$\pm$0.7 & $>$365 \\  %% 13.2  5.1
 12&    57.8$\pm$0.8 & (2.0) &        --     &        --     &       --     &   2.3$\pm$1.0 &      --     &    --  \\  %%
\hline
\multicolumn{1}{c}{ } & \multicolumn{2}{l}{10--20 km~s$^{-1}$} &
                              488.7$\pm$15. & 169.9$\pm$6.2 & 35.9$\pm$2.8 & 200.2$\pm$7.9 & 10.4$\pm$1.0 &     40 \\  %% 19.3  8.5
\multicolumn{1}{c}{ } & \multicolumn{2}{l}{outliers} &
                            $<$71.0         &  98.1$\pm$5.1 &$<$36.6       & 286.8$\pm$7.7 & 19.5$\pm$1.7 & $>$405 \\  %% 14.7  3.4
\multicolumn{1}{c}{ } & \multicolumn{2}{l}{total}    & 
                             (526.0$\pm$18.)& 268.0$\pm$8.0 &(63.0$\pm$5.0)& 487.0$\pm$11. & 29.9$\pm$2.0 &     95 \\  %% 16.3  5.5
\hline
\end{tabular}
\medskip
~ ~ \\
Total column densities for \mbox{Zn\,{\sc ii}} and \mbox{Cr\,{\sc ii}} in parentheses should be considered as lower limits, as not all components were detected.
\end{minipage}
\end{table*}

The 1~$\sigma$ uncertainties given for $N$, $b$, and $v$ in Tables~\ref{tab:comps} and \ref{tab:compuv} reflect only photon noise, and thus should be considered as lower limits to the true uncertainties.
Consideration of the effects of uncertainties in continuum placement (most significant for the stronger lines) and of allowing fixed parameters (usually $b$) to vary would increase the uncertainties in the various parameters (primarily $N$).
In view of the statistics of component width and separation obtained from higher resolution, higher S/N spectra of Galactic targets (Welty et al. 1994, 1996; Welty \& Hobbs 2001), however, it is likely that the `true' component structures are even more complex than those listed in the tables -- and that uncertainty may well be even more significant for the individual component parameters.
Differences in the component velocities derived for \mbox{Na\,{\sc i}} and \mbox{Ca\,{\sc ii}}, for example, may be due to component-to-component differences in relative abundances in an underlying more complex structure.
Higher resolution spectra of \mbox{Ca\,{\sc ii}} toward HD~72127A do exhibit more complex structure (Fig.~\ref{fig:naca}; Welty et al. 1996).
Differences between the component velocities derived from optical and UV spectra of HD~72127A may reflect (in addition) possible temporal variations between 1992 and 2003.

The total sight line column densities are given in Table~\ref{tab:coldens}, with values for the main neutral components toward 23 Ori (Welty et al. 1999b) and $\zeta$ Oph (Welty et al. 2003 and references therein) given for comparison.
In all cases, the values derived from the profile fits are consistent with estimates or limits based on integrating the `apparent' optical depth over the absorption lines; the total sight line column densities are in most cases fairly well determined.

\begin{table}
\caption{Column densities.}
\label{tab:coldens} 
\begin{tabular}{@{}lrrrr}
\hline
\multicolumn{1}{l}{Species}&
\multicolumn{1}{c}{HD~72127A}& 
\multicolumn{1}{c}{HD~72127B}&
\multicolumn{1}{c}{23 Ori SLV}&
\multicolumn{1}{c}{$\zeta$ Oph B}\\  
\hline
%%%  Species             72127 A          72127 B       23 Ori S  zet Oph B
H$_{\rm tot}$       &[20.40]         &[20.40]         &    20.71 &    21.15 \\
\mbox{O\,{\sc i}}   & $<$17.14       &     --         &    17.29 &    17.63 \\
\mbox{Na\,{\sc i}}  & 12.46$\pm$0.12 & 12.66$\pm$0.06 &    13.41 &    13.85 \\
\mbox{Mg\,{\sc i}}  & 12.92$\pm$0.04 &     --         &    13.80 &    13.80 \\
\mbox{Mg\,{\sc ii}} & 15.29$\pm$0.04 &     --         &    15.57 &    15.32 \\
\mbox{Si\,{\sc i}}  & 11.20$\pm$0.11 &     --         &    11.79 &     --   \\
\mbox{Si\,{\sc ii}} & 15.43$\pm$0.01 &     --         &    15.37 &    15.34 \\
\mbox{Al\,{\sc i}}  & $<$10.83       & $<$10.83       & $<$11.53 & $<$10.22 \\
\mbox{Cl\,{\sc i}}  & 12.52$\pm$0.16 &     --         &    13.29 &    14.48 \\
\mbox{Ca\,{\sc i}}  &  9.58$\pm$0.08 & 10.76$\pm$0.02 &    10.20 &     9.46 \\
\mbox{Ca\,{\sc ii}} & 12.99$\pm$0.01 & 13.22$\pm$0.02 &    12.04 &    11.71 \\
\mbox{Ti\,{\sc i}}  & $<$10.88       & $<$10.79       & $<$10.72 &     --   \\
\mbox{Ti\,{\sc ii}} & 11.68$\pm$0.06 & 11.78$\pm$0.08 &    11.15 &    11.03 \\
\mbox{Cr\,{\sc i}}  & $<$10.61       & 11.26$\pm$0.02 & $<$10.59 &     9.78 \\
\mbox{Cr\,{\sc ii}} & 12.80$\pm$0.04 &     --         &    12.64 &    12.39 \\
\mbox{Mn\,{\sc i}}  & $<$11.17       & $<$11.17       &    10.23 & $<$11.30 \\
\mbox{Mn\,{\sc ii}} &     --         &     --         &    13.08 &    13.20 \\
\mbox{Fe\,{\sc i}}  & 10.77$\pm$0.11 & 11.89$\pm$0.03 &    11.34 &    11.23 \\
\mbox{Fe\,{\sc ii}} & 14.69$\pm$0.01 &     --         &    14.30 &    14.32 \\
\mbox{Ni\,{\sc i}}  & $<$11.95       & $<$12.03       & $<$10.26 & $<$11.40 \\
\mbox{Ni\,{\sc ii}} & 13.48$\pm$0.03 &     --         &    12.90 &    12.91 \\
\mbox{Zn\,{\sc ii}} & 12.72$\pm$0.01 &     --         &    13.24 &     --   \\
\mbox{Ge\,{\sc ii}} & 11.25$\pm$0.17 &     --         &    11.86 &    11.75 \\
\hline
\end{tabular}
\medskip
~ ~ \\
Entries are log [$N$ (cm$^{-2}$)] $\pm$ 1$\sigma$ uncertainty, based on fits to the line profiles; limits are 3$\sigma$. 
Values for 23 Ori SLV (main) components are from Welty et al. (1999b), except for \mbox{Na\,{\sc i}}, \mbox{Ti\,{\sc i}}, and \mbox{Cr\,{\sc i}}, which are from new UVES data. 
Values for $\zeta$ Oph B (main) component are from Welty \& Hobbs (2001), Welty et al. (2003), and references therein.
\end{table}

%%%%%%%%%%%%%%%%%%%%%%%%%%%%%%%%%% DISCUSSION %%%%%%%%%%%%%%%%%%%%%%%%%%%%%%%%

\section{Discussion}
\label{sec-disc}

\subsection{Abundances and depletions}
\label{sec-abund}

In view of the spectral types of HD~72127A and HD~72127B, it is likely that the Lyman-$\alpha$ absorption observed toward the two stars (in archival {\it IUE} spectra) is an inseparable blend of stellar and interstellar components.
Several indirect methods thus have been used to estimate the interstellar hydrogen column densities toward the two stars: \\
(1) The first makes use of the strong, nearly linear correlation observed between the equivalent width of the diffuse interstellar band (DIB) at 5780 \AA\ and $N$(\mbox{H\,{\sc i}}) (e.g., Herbig 1993, 1995; Welty et al. 2006; D. G. York et al., in preparation).
The equivalent widths estimated from our UVES spectra -- $W$(5780) = 15$\pm$6 m\AA\ for HD~72127A and $W$(5780) = 12$\pm$5 m\AA\ for HD~72127B -- suggest log[$N$(\mbox{H\,{\sc i}})] of about 20.2 for HD~72127A and about 20.1 for HD~72127B.
The observed $W$(5780) are somewhat low, relative to the corresponding $N$(\mbox{Na\,{\sc i}}), however, suggesting that the 5780 \AA\ DIB may be weaker than usual toward HD~72127AB -- as has been observed, for example, toward some Galactic and Magellanic Clouds targets (Herbig 1993; Welty et al. 2006).
The $N$(\mbox{H\,{\sc i}}) estimated from $W$(5780) might thus be best viewed as lower limits. \\
(2) An estimate for $N$(H) toward HD~72127A may be obtained from the observed column density of \mbox{Zn\,{\sc ii}}, as zinc is generally only very mildly depleted in diffuse interstellar clouds.
If the depletion of zinc D(Zn) = $-$0.2 dex (typical for warm diffuse gas), then log[$N$(\mbox{Zn\,{\sc ii}})] = 12.72 implies that log[$N$(H)] would be about 20.3, for a solar zinc abundance of $-$7.37 dex (Lodders 2003). 
Fits to the lower resolution G160M spectra of the \mbox{S\,{\sc ii}} lines at 1250, 1253, and 1259 \AA\ using the component structures found for \mbox{Zn\,{\sc ii}} and \mbox{Si\,{\sc ii}} yield consistent estimates for $N$(H), assuming a solar abundance of $-$4.81 dex and no depletion for sulphur.\\
(3) If the typical Galactic gas-to-dust ratio log[$N$(H)/$E(B-V)$] = 21.71 (e.g., Welty et al. 2006) applies to the sight line toward HD~72127AB, then the observed $E(B-V)$ = 0.10 would imply log[$N$(H)] $\sim$ 20.7$\pm$0.3 toward both stars. \\
(4) The upper limit on $N$(\mbox{O\,{\sc i}}) (from the non-detection of the weak $\lambda$1355 line) toward HD~72127A may be used to place a corresponding upper limit on $N$(\mbox{H\,{\sc i}}), as the two species are strongly coupled via charge exchange.
The observed log[$N$(\mbox{O\,{\sc i}})] $<$ 17.14 implies log[$N$(\mbox{H\,{\sc i}})] $<$ 20.55, for a solar oxygen abundance of $-$3.31 dex and a typical `warm cloud' depletion of $-$0.1 dex (e.g., Cartledge et al. 2004). \\
(5) The equivalent widths observed for the weak lines of typically moderately depleted \mbox{Mg\,{\sc ii}} (1239 and 1240 \AA) and for the strong lines of typically undepleted \mbox{S\,{\sc ii}} (1250, 1253, and 1259 \AA) are very similar toward the two stars (Table~\ref{tab:ewids}; Wallerstein et al. 1995b) -- suggesting that $N$(H) also is similar toward both. \\
(6) The upper limit on log[$N$(CH)] $<$ 12.4 obtained from the UVES spectra suggests that log[$N$(H$_2$)] is less than 20.0 toward both stars (e.g., Welty et al. 2006). \\
As the different indicators (photometry, optical spectra, UV spectra) are from different epochs, these comparisons and estimates are somewhat uncertain, given the observed temporal variations for some interstellar quantities in both sight lines.
None the less, in view of the various estimates and limits, we have adopted log[$N$(\mbox{H\,{\sc i}})] $\sim$ log[$N$(H)] $\sim$ 20.4, with an uncertainty of perhaps $\pm$50 percent, toward both HD~72127A and HD~72127B.

\begin{table}
\caption{Abundance and depletion data (HD~72127A).}
\label{tab:depl}
\begin{tabular}{@{}lrcccc}
\hline
\multicolumn{1}{l}{Element}&
\multicolumn{1}{c}{Solar}&
\multicolumn{1}{c}{D(Warm)/}&
\multicolumn{3}{c}{D(72127A)}\\
\multicolumn{1}{c}{ }&
\multicolumn{1}{c}{ }&
\multicolumn{1}{c}{D(Cold)}&
\multicolumn{1}{c}{main}&
\multicolumn{1}{c}{outlier}&
\multicolumn{1}{c}{total}\\
\hline
%%%  solar      warm/cold    D(main)    D(out)   D(total)
%H  & 12.00 &      --/--    &  [20.32]&  [19.60]&  [20.40]\\
O  &  8.69 & $-$0.1/$-$0.2 & $<$0.03 &    --   &$<-$0.05 \\
Na &  6.30 & $-$0.6/$-$0.6 &   --    &    --   &    --   \\
Mg &  7.55 & $-$0.6/$-$1.3 &   --    &    --   & $-$0.66 \\
Si &  7.54 & $-$0.4/$-$1.3 & $-$0.63 & $-$0.15 & $-$0.51 \\
Ca &  6.34 & $-$2.0/$-$3.6 & $-$2.38 &    --   & $-$1.75 \\
Ti &  4.92 & $-$1.3/$-$2.9 & $-$1.77 & $-$1.25 & $-$1.64 \\
Cr &  5.65 & $-$1.2/$-$2.2 & $-$1.41 & $-$0.82 & $-$1.25 \\
Fe &  7.47 & $-$1.3/$-$2.2 & $-$1.49 & $-$0.61 & $-$1.18 \\
Ni &  6.22 & $-$1.4/$-$2.3 & $-$1.52 & $-$0.53 & $-$1.14 \\
Zn &  4.63 & $-$0.2/$-$0.4 & $-$0.26 &    --   & $-$0.31 \\
Ge &  3.62 & $-$0.6/$-$0.9 & $-$0.69 &    --   &    --   \\
\hline
\end{tabular}
\medskip
~ ~ \\
Solar abundances are from Lodders (2003).
Representative warm and cold cloud depletions are updates of values in Welty et al. (1999b).
Last three columns give depletions for HD~72127A:  main components (10--20 km~s$^{-1}$), all other components, and all components, respectively. 
Those depletions assume log[$N$(H)] = 20.32, 19.60, and 20.40 dex for the three groups. 
\end{table}

The fits to the \mbox{Zn\,{\sc ii}}, \mbox{S\,{\sc ii}}, and \mbox{Si\,{\sc ii}} profiles toward HD~72127A suggest that roughly 80--90 percent of the total \mbox{H\,{\sc i}} is found in the components between 10 and 20 km~s$^{-1}$ (`main components'), with the remainder in the outlying higher velocity components.
On the other hand, roughly 60 percent of the \mbox{Fe\,{\sc ii}} and \mbox{Ni\,{\sc ii}} is found in the outlying components.
Inspection of the ratios of the column densities of typically depleted species to that of (nearly) undepleted \mbox{Zn\,{\sc ii}} (e.g., \mbox{Fe\,{\sc ii}}/\mbox{Zn\,{\sc ii}}; last column of Table~\ref{tab:compuv}) indicates that the depletions in the outlying components are significantly less severe than those in the main components.
If the adopted total $N$(H) is apportioned to the main components (2.1 $\times$ 10$^{20}$ cm$^{-2}$) and the outlying components (4 $\times$ 10$^{19}$ cm$^{-2}$), then the depletions may be estimated separately for those two component groups (Table~\ref{tab:depl}).
For the main components, the resulting depletions are broadly consistent (within the uncertainties) with those in `warm diffuse' clouds in the Galactic disc (Savage \& Sembach 1996; Welty et al. 1999); for the outlying components, the depletions are more similar to those in Galactic halo clouds.
The similarities in $E(B-V)$, $N$(\mbox{Ti\,{\sc ii}}), and the equivalent widths of the 5780 \AA\ DIB and the lines of \mbox{Mg\,{\sc ii}} and \mbox{S\,{\sc ii}} for both HD~72127A and HD~72127B (Wallerstein et al. 1995b; this paper) suggest that both the total $N$(H) and the average depletions are also similar (within factors of about 2) for the two lines of sight; the depletions may be slightly less severe toward HD~72127B.

Examination of the profiles of the \mbox{Al\,{\sc iii}} lines at 1854 and 1862 \AA\ may provide some insight into the distribution of ionized gas toward HD~72127A.
Integrating the apparent optical depths over the profiles yields a total $N$(\mbox{Al\,{\sc iii}}) of about 1.7$\times$10$^{13}$ cm$^{-2}$ in each case, so the lines do not appear to be significantly saturated. 
There is very little \mbox{Al\,{\sc iii}} absorption at the velocities of the main components between 10 and 20 km~s$^{-1}$ -- which thus are likely to be predominantly neutral.
The \mbox{Al\,{\sc iii}} absorption at higher and lower velocities (especially the strong absorption centred at about 35 km~s$^{-1}$) signals the presence of some ionized gas at those outlying velocities.
Crude fits to the \mbox{Al\,{\sc ii}} line at 1670 \AA\ (using the \mbox{Fe\,{\sc ii}} component structure), however, suggest that the \mbox{Al\,{\sc ii}}/\mbox{Al\,{\sc iii}} ratio is about 3--4 (on average) for those outlying components -- so that the fraction of ionized gas may be relatively small there as well.
Uncertainties as to the amounts of \mbox{H\,{\sc ii}} and \mbox{X\,{\sc iii}} (i.e., the second ions of elements whose first ions \mbox{X\,{\sc ii}} are dominant in \mbox{H\,{\sc i}} gas) present in the outlying components imply corresponding uncertainties in the depletions estimated for those components in Table~\ref{tab:depl}. 

\subsection{\mbox{Ca\,{\sc ii}}/\mbox{Na\,{\sc i}} ratios}
\label{sec-naca}

The \mbox{Ca\,{\sc ii}}/\mbox{Na\,{\sc i}} ratio is usually considered to be an indicator of the (highly variable) depletion of calcium, and can range from $\la$0.01 in cold, dense clouds (where calcium generally is severely depleted) to $\ga$10 in warm, diffuse gas (where some fraction of the calcium has been returned to the gas phase).
Hobbs et al. (1982) found unusually high \mbox{Ca\,{\sc ii}}/\mbox{Na\,{\sc i}} ratios ($>$9) for all five of the components toward HD~72127A seen in 1977.
The detailed fits reported by Hobbs et al. (1991), however, yielded average ratios ranging from 1.4 for the component at $+$13 km~s$^{-1}$ (with strongest \mbox{Na\,{\sc i}}) to $>$30 for the components at $-$19 and $-$8 km~s$^{-1}$, for observations between 1981 and 1988. 
Fits to the 2003 UVES spectra imply \mbox{Ca\,{\sc ii}}/\mbox{Na\,{\sc i}} ratios ranging from 0.6 (at $+$13 km~s$^{-1}$) to $>$20 (at $-$15, $+$40, and $+$55 km~s$^{-1}$) toward HD~72127A and from 2.7 (at $+$11 km~s$^{-1}$) to $>$40 (at $-$16 and $+$54 km~s$^{-1}$) toward HD~72127B.
At least some of the differences between the current and former results are likely due to the more detailed and precise component information obtained from the more recent spectra and to the availability of $\lambda$3302 data for \mbox{Na\,{\sc i}}.
The depletions that would be inferred from the \mbox{Ca\,{\sc ii}}/\mbox{Na\,{\sc i}} ratios in the different components thus are consistent with the depletion levels determined more directly from the various dominant species in the previous section.

\begin{table}
\caption{\mbox{Ca\,{\sc ii}} component velocities (1988--2003).}
\label{tab:cavel}
\begin{tabular}{@{}rrrrr}
\hline
\multicolumn{1}{c}{1988 Dec}&
\multicolumn{1}{c}{1994 Feb}&
\multicolumn{1}{c}{1994 Feb}&
\multicolumn{1}{c}{1996 J/F}&
\multicolumn{1}{c}{2003 Nov}\\
\hline
\multicolumn{5}{c}{HD~72127A}\\
\hline
        &         &       &       & $-$27.7 \\
$-$18.8 & $-$17.7 &       &       &         \\
$-$14.6 & $-$15.3 & $-$15 & $-$15 & $-$15.4 \\
        &  $-$9.1 &       &       &         \\
 $-$8.0 &  $-$8.2 &  $-$8 &  $-$8 &  $-$8.2 \\
        &  $-$0.8 &       &       &         \\
        &     2.0 &     3 &     3 &     2.9 \\
    3.5 &     4.6 &       &       &         \\
        &     7.8 &       &       &         \\
   11.0 &    12.7 &       &       &    12.6 \\
        &    13.2 &       &       &         \\
   17.4 &    14.3 &    14 &    14 &         \\
        &    22.1 &       &       &    18.3 \\
   27.7 &    29.8 &    30 &    29 &    26.0 \\
        &    31.1 &       &       &    32.3 \\
        &    38.5 &       &       &         \\
   39.8 &    40.9 &    41 &    40 &    40.0 \\
   54.8 &    53.9 &    55 &    55 &    54.7 \\
   59.8 &    58.2 &       &       &         \\
\hline
\multicolumn{5}{c}{HD~72127B}\\
\hline
        &         &       & $-$15 & $-$16.0 \\
        &         &       &  $-$6 &  $-$5.4 \\
        &         &       &       &     4.3 \\
        &         &       &    13 &    10.5 \\
        &         &       &       &    21.0 \\
        &         &       &    30 &    29.4 \\
        &         &       &       &    35.7 \\
        &         &       &    40 &    42.9 \\
        &         &       &    55 &    53.8 \\
        &         &       &       &    88.0 \\
        &         &       &       &    96.9 \\
\hline
\end{tabular}
\medskip
~ ~ \\
Velocities are heliocentric; ultra-high resolution spectrum from 1994 Feb used as standard for HD~72127A.
References:  1988 Dec (Hobbs et al. 1991); 1994 Feb (Welty et al. 1996); 1994 Feb and 1996 J/F (Cha \& Sembach 2000); 2003 Nov (this paper).
See Hobbs et al. (1982, 1991) for values from previous epochs.
\end{table}
 
\subsection{Temporal variations}
\label{sec-tvar}

Temporal variations over the past 30 years in the complex profiles of \mbox{Na\,{\sc i}} and \mbox{Ca\,{\sc ii}} toward HD~72127A have been well documented.
Hobbs et al. (1982) noted the appearance of a `new' component at $+$15 km~s$^{-1}$ in a \mbox{Ca\,{\sc ii}} spectrum obtained in 1981 April.
Detailed fits to that same spectrum (and others obtained through 1988 Dec) revealed both additional weaker components and variations in the column densities and/or velocities of most of the \mbox{Ca\,{\sc ii}} components between $-$15 and $+$17 km~s$^{-1}$ (Hobbs et al. 1991).
An ultra-high resolution \mbox{Ca\,{\sc ii}} spectrum obtained in 1994 Feb revealed continued variations over that velocity range, more complex structure between $-$20 and $+$40 km~s$^{-1}$, and new variations between $+$25 and $+$35 km~s$^{-1}$ (Fig.~\ref{fig:naca}; Welty et al. 1996; see also Cha \& Sembach 2000).
The UVES \mbox{Ca\,{\sc ii}} spectrum (2003 Nov) shown in Fig.~\ref{fig:naca} indicates further changes in the component column densities over most of the velocity range and continued evolution of the structure between $+$25 and $+$35 km~s$^{-1}$.
The component velocities found for the UVES spectrum are very similar, however, to those determined from spectra of comparable resolution obtained in 1994 and 1996 (Cha \& Sembach 2000), when allowance is made for differences in adopted component structure (Table~\ref{tab:cavel}).
The most noticeable change in the \mbox{Na\,{\sc i}} profile toward HD~72127A since 1988 is the appearance of a distinct component near $+$4 km~s$^{-1}$ (Cha \& Sembach 2000; this paper) superimposed on (or in place of) the smooth, broad shoulder seen in previous (higher resolution) spectra short-ward of the strongest \mbox{Na\,{\sc i}} absorption near $+$13 km~s$^{-1}$.

Unfortunately, much less information is available for the interstellar lines toward the fainter HD~72127B.
Component velocities for \mbox{Ca\,{\sc ii}} reported by Hobbs et al. (1982) and by Cha \& Sembach (2000) are reasonably consistent with those determined from the UVES spectrum (Table~\ref{tab:cavel}), and the 1996 line profile plotted by Cha \& Sembach (2000) appears to be very similar to the 2003 UVES profile plotted in Fig.~\ref{fig:naca} (though the total equivalent width was smaller in 2003).
The weak high-velocity components found in the UVES spectrum of HD~72127B (similar to those seen toward several other stars in the Vela region; e.g., Wallerstein et al. 1980; Danks \& Sembach 1995) do not appear to be present in the 1996 spectrum, however.
The lines from several of the trace neutral species measured in the 2003 UVES spectrum of HD~72127B are significantly stronger than previously reported:  the equivalent width of the \mbox{Na\,{\sc i}} $\lambda$5895 line (188 m\AA) is roughly twice the 95 m\AA\ value listed by Hobbs et al. (1982), while the equivalent width of the \mbox{Ca\,{\sc i}} $\lambda$4226 line (16 m\AA) is more than three times the upper limit given by Wallerstein \& Gilroy (1992).

\subsection{Spatial variations}
\label{sec-spvar}

Thackeray (1974) first called attention to differences in the profiles of the strong, broad \mbox{Ca\,{\sc ii}} K lines toward HD~72127A and HD~72127B; Hobbs et al. (1982) subsequently noted differences in the velocity structures seen in both \mbox{Ca\,{\sc ii}} and \mbox{Na\,{\sc i}}.
Higher resolution \mbox{Ca\,{\sc ii}} spectra obtained by Cha \& Sembach (2000) indicate that while the overall velocity extent of the absorption is very similar toward the two stars, the detailed column density distribution (as a function of velocity) is quite different.
Wallerstein et al. (1995b) found similar equivalent widths for UV lines of \mbox{Mg\,{\sc ii}} and \mbox{S\,{\sc ii}} (but differences in the strengths of lines from \mbox{P\,{\sc ii}} and \mbox{Ge\,{\sc ii}}) in {\it HST}/G160M spectra of the two stars.
Possible differences in detailed component structure (as seen for \mbox{Ca\,{\sc ii}}) could not be discerned in those moderate-resolution UV spectra, however.

The UVES spectra of HD~72127A and HD~72127B shown in Figs.~\ref{fig:naca} and \ref{fig:weak} provide both a more recent snapshot of the differences in \mbox{Na\,{\sc i}} and \mbox{Ca\,{\sc ii}} and indications of similarities and differences for lines from several other species.
Toward HD~72127B, the main component(s) near $+$11 to $+$13 km~s$^{-1}$ are stronger in both \mbox{Na\,{\sc i}} and \mbox{Ca\,{\sc ii}} than those toward HD~72127A, but the outlying components generally are weaker. 
The overall structure and relative column densities are more similar for \mbox{Na\,{\sc i}} than for \mbox{Ca\,{\sc ii}} -- opposite the general trend for observations of those two species toward other binary systems (Meyer 1990; Watson \& Meyer 1996).
The weak \mbox{Na\,{\sc i}} $\lambda$3302 doublet lines are similar in strength toward the two stars, but there are modest differences in the profiles of the \mbox{Ti\,{\sc ii}} $\lambda$3383 line and very striking differences in the lines from previously undetected \mbox{Ca\,{\sc i}}, \mbox{Cr\,{\sc i}}, and \mbox{Fe\,{\sc i}} (which are much stronger toward HD~72127B; see next section).

\subsection{Trace neutral species toward HD~72127B}
\label{sec-trace}

The absorption lines from several trace neutral species are unusually strong in the 2003 UVES spectrum of HD~72127B.
For the $N$(H) $\sim$ 2.5$\times$10$^{20}$ cm$^{-2}$ estimated above, the column density of \mbox{Na\,{\sc i}} is somewhat higher than usual (Welty \& Hobbs 2001), and the column densities of \mbox{Ca\,{\sc i}}, \mbox{Cr\,{\sc i}}, and \mbox{Fe\,{\sc i}} are remarkably high (Welty et al. 2003) -- especially as most of the absorption from those trace neutral species is found in the (apparently) single strong component near $+$11 km~s$^{-1}$.
The column density of \mbox{Ca\,{\sc i}}, for example, is among the highest known for any Galactic sight line; comparable values generally are seen only for sight lines with much higher $E(B-V)$ and/or $N$(H).
Absorption from interstellar \mbox{Cr\,{\sc i}} had previously been seen only toward $\zeta$ Oph (Meyer \& Roth 1990), with an equivalent width $W$(3578) = 0.25$\pm$0.06 m\AA\ for the strongest line at 3578 \AA\ in the main (B) components near $-$15 km~s$^{-1}$, which have $N$(H) $\sim$ 14$\times$10$^{20}$ cm$^{-2}$.
Toward HD~72127B, the \mbox{Cr\,{\sc i}} $\lambda$3578 line has $W$(3578) = 5.4$\pm$0.7 m\AA, and five other \mbox{Cr\,{\sc i}} lines are seen at greater than 3$\sigma$ significance -- all at the same velocity and with equivalent widths consistent with the relative $f$-values (Table~\ref{tab:ewids}).

\begin{table}
\caption{Ratios of trace neutral species (predicted and observed).}
\label{tab:ratios}
\begin{tabular}{@{}lrrrr}
\hline
\multicolumn{1}{l}{Location}&
\multicolumn{1}{c}{\mbox{Mg\,{\sc i}}/\mbox{Na\,{\sc i}}}&
\multicolumn{1}{c}{\mbox{Ca\,{\sc i}}/\mbox{Na\,{\sc i}}}&
\multicolumn{1}{c}{\mbox{Cr\,{\sc i}}/\mbox{Na\,{\sc i}}}&
\multicolumn{1}{c}{\mbox{Fe\,{\sc i}}/\mbox{Na\,{\sc i}}}\\
\hline
%%%                     Mg I      Ca I      Cr I      Fe I
Solar (7000 K)      & $+$1.55 & $+$0.08 & $-$2.43 & $+$0.92 \\
Solar (3000 K)      & $+$0.52 & $-$1.65 & $-$2.43 & $+$0.02 \\
Solar (100 K)       & $+$0.60 & $-$1.29 & $-$2.43 & $+$0.10 \\
%%%
Warm (7000 K)       & $+$1.55 & $-$1.32 & $-$3.03 & $+$0.22 \\
Warm (3000 K)       & $+$0.52 & $-$3.05 & $-$3.03 & $-$0.68 \\
%%%
Cold (100 K)        & $-$0.10 & $-$4.29 & $-$4.03 & $-$1.50 \\  %%% rad recomb dominates all
%%%                 & $-$0.03 & $-$4.45 &    --   & $-$1.56 \\  %%% grain assisted recomb dominates all (s=0; old base values)
 ~ ~ \\
HD~72127A           & $+$0.46 & $-$2.88 &$<-$1.85 & $-$1.69 \\  %%% W30/S1/S30 %%% W30        %%% all    %%% C1     %%% depl vs. b(Ca I)
HD~72127B           &    --   & $-$1.90 & $-$1.46 & $-$0.79 \\  %%%            %%% S30/W50    %%% S70    %%% W30
23 Ori SLV          & $+$0.39 & $-$3.21 &$<-$2.82 & $-$2.07 \\  %%% W30/S1/S30 %%% W30        %%% W30/C1 %%% <C1
$\zeta$ Oph B       & $-$0.05 & $-$4.39 & $-$4.05 & $-$2.62 \\  %%% C1         %%% C1         %%% C1     %%% <C1
 ~ ~ \\
$\epsilon$ Ori (+3) & $+$0.81 & $-$1.77 &    --   &    --   \\
HD~90177 ($-$40)    &    --   & $-$1.37 &    --   & $+$0.36 \\  %%%            %%% W70/S1/S50 %%%        %%% S1/S50
HD~94910 ($-$54)    &    --   & $-$1.14 &    --   & $+$0.54 \\  %%%            %%% W70/S1/S50 %%%        %%% S50
Sk$-$67~5 (+261)    &    --   &$>-$0.85 &    --   &$>+$1.05 \\  %%%            %%% S50/S70    %%%        %%% S70
Sk$-$68~52 (+307)   &    --   & $-$1.51 &    --   &$<+$0.31 \\
SN1987A  (+65)      &    --   & $-$1.57 &    --   &    --   \\  %%%            %%% W60/S30/S1 %%%        %%%
SN1987A (+165)      &    --   &$>-$0.83 &    --   &    --   \\  %%%            %%% S30/S70    %%%        %%%
SN1987A (+216)      &    --   & $-$1.74 &    --   &    --   \\  %%%            %%% W50/S30    %%%        %%%
\hline
\end{tabular}
\medskip
~ ~ \\
Entries are log [$N$(X)/$N$(Y)]; limits are 3$\sigma$. 
Predicted ratios for solar, warm cloud, and cold cloud abundances use the solar abundances and representative depletions from Table~5 and ionization equilibrium calculations as in eqn.~1 (see also Welty et al. 1999a).
Predicted values for \mbox{Ca\,{\sc i}} are for $n_e$ = 1.0 cm$^{-3}$.
Values for HD~90177 and HD~94910 are from Gnaci\'{n}ski \& Krogulec (2008); values for Sk$-$67~5 and Sk$-$68~52 are from Welty \& Crowther (in preparation); values for SN1987A are from Welty et al. (1999a).
\end{table}

Such (relatively) strong absorption from \mbox{Ca\,{\sc i}} and/or \mbox{Fe\,{\sc i}} has recently been noted in a few other sight lines.
Table~\ref{tab:ratios} compares the ratios of several trace neutral species \mbox{X\,{\sc i}}/\mbox{Na\,{\sc i}} -- as observed toward HD~72127A, HD~72127B, and several other targets -- with the values that would be predicted assuming ionization equilibrium under various conditions.
The last six sight lines in the table exhibit components with even more extreme values of \mbox{Ca\,{\sc i}}/\mbox{Na\,{\sc i}} and/or \mbox{Fe\,{\sc i}}/\mbox{Na\,{\sc i}}; the well-studied sight lines toward 23~Ori (Welty et al. 1999b) and $\zeta$~Oph (e.g., Welty \& Hobbs 2001; Welty et al. 2003) are included for comparison.
The clouds toward HD~90177 and HD~94910 are two of the so-called `CaFe' clouds identified by Bondar et al. (2007) and discussed by Gnaci\'{n}ski \& Krogulec (2008).
Still higher values of \mbox{Ca\,{\sc i}}/\mbox{Na\,{\sc i}} and/or \mbox{Fe\,{\sc i}}/\mbox{Na\,{\sc i}} are seen in components toward three targets in the Large Magellanic Cloud (Welty et al. 1999a; Welty \& Crowther, in preparation).
In two of those LMC components, no absorption is detected in the \mbox{Na\,{\sc i}} D lines, with upper limits on $N$(\mbox{Na\,{\sc i}}) of order 1--4$\times$10$^{10}$ cm$^{-2}$.

Observations of quasar absorption-line systems have begun to yield detections of weak lines from various trace neutral species (besides the commonly observed \mbox{Mg\,{\sc i}}) in higher redshift systems.
For example, \mbox{Si\,{\sc i}}, \mbox{Fe\,{\sc i}}, and several other neutral species have been detected (along with a modest amount of H$_2$) in a sub-Damped Lyman $\alpha$ system at $z$ $\sim$ 1.15 (Quast, Reimers, \& Baade 2008).
Unexpectedly strong absorption from \mbox{Si\,{\sc i}}, \mbox{Ca\,{\sc i}}, and \mbox{Fe\,{\sc i}} has also been observed in a weak \mbox{Mg\,{\sc ii}} absorber at $z$ $\sim$ 0.45 (D'Odorico 2007)\footnotemark.
Absorption from \mbox{Na\,{\sc i}} has not yet been observed in those systems, however.
\footnotetext{Based on comparisons with Galactic sight lines, D'Odorico (2007) infers log[$N$(H)] $\sim$ 20.5--21.5 -- and thus a very low metallicity -- for that absorber.
The Galactic sight lines, however, are typically characterized by significant depletions of the various refractory elements.
The $N$(H) in the weak \mbox{Mg\,{\sc ii}} system (where the depletions are likely much less severe) may well be much lower, and the metallicity correspondingly higher.} 

Consideration of the physical conditions and ionization equilibrium in the main cloud toward HD~72127B might provide some insight into the high observed abundances of the trace neutral species in those various contexts.
Wallerstein et al. (1995b) found a relatively high fraction of the \mbox{C\,{\sc i}} toward HD~72127B to be in the excited fine-structure states -- indicative of a relatively high thermal pressure ($n_{\rm H}T$) in the gas.
High densities would be expected to favor the trace neutral species; \mbox{Ca\,{\sc i}} (and several other neutral species) can be enhanced via dielectronic recombination in warmer gas.
While the width of the \mbox{Ca\,{\sc i}} component at $+$13 km~s$^{-1}$ toward HD~72127A ($b$ = 0.6 km~s$^{-1}$; Welty et al. 2003) places an upper limit on the temperature of $T$ $\la$ 850 K for that component (so that dielectronic recombination would not be significant there), the main \mbox{Ca\,{\sc i}} component toward HD~72127B may be somewhat broader.
Higher resolution spectra of HD~72127B would be needed to place a firmer limit on the temperature (and the role of dielectronic recombination) in the main component in that sight line.
The \mbox{Ca\,{\sc i}}/\mbox{Ca\,{\sc ii}} ratio (0.0055) in the main component toward HD~72127B, however, is very similar to the ratios seen in other Galactic sight lines (e.g., Welty et al. 2003) and is only a factor of two higher than the ratio (0.0026) found in the main component toward HD~72127A -- which would seem to suggest that \mbox{Ca\,{\sc i}} is not greatly enhanced relative to \mbox{Ca\,{\sc ii}} and that the electron density $n_e$ is not especially high.

\begin{figure}
\includegraphics[width=84mm]{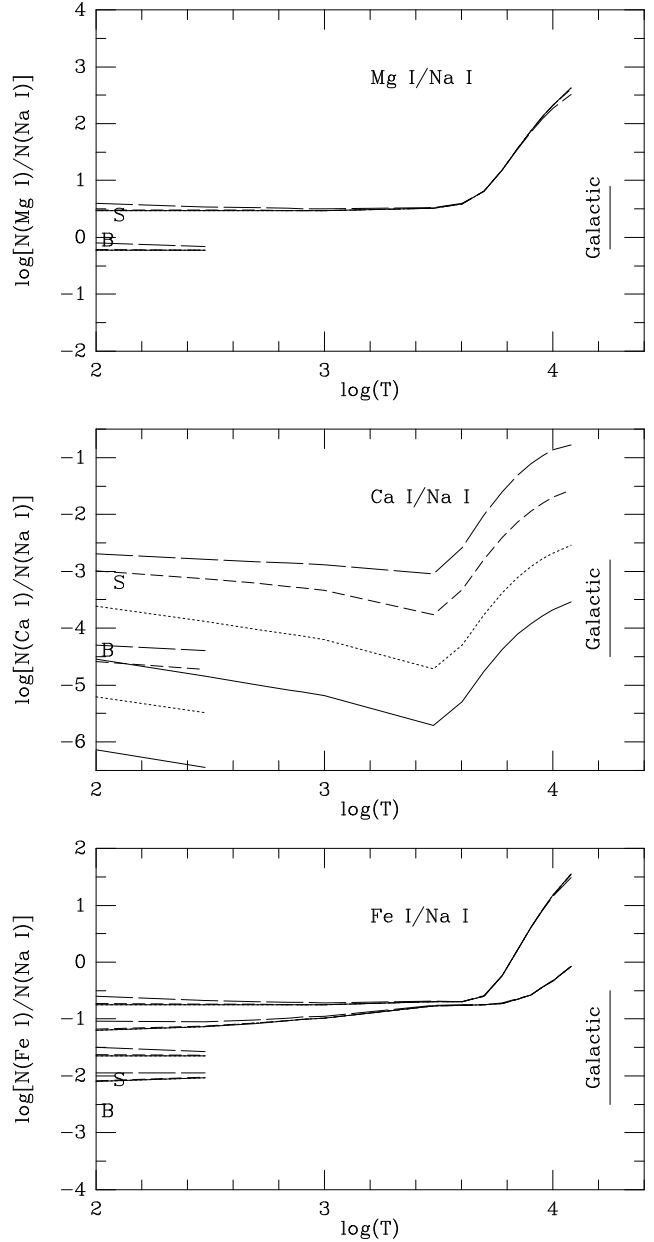}
\caption{Predicted ratios \mbox{X\,{\sc i}}/\mbox{Na\,{\sc i}} -- for X = Mg, Ca, and Fe -- as functions of temperature.
The calculations assume equilibrium between ionization (radiative and collisional) and recombination (radiative and dielectronic), warm cloud depletions, and $n_e$ = 0.001, 0.01, 0.1, or 1.0 cm$^{-3}$ (lower to upper curves in each panel).
The lower partial curves (from 100 to 300 K) show the ratios for cold cloud depletions.
The increases at higher temperatures are due to dielectronic recombination to \mbox{Mg\,{\sc i}}, \mbox{Ca\,{\sc i}}, and \mbox{Fe\,{\sc i}}.
The lower set of curves for \mbox{Fe\,{\sc i}} shows the predictions for the photoionization and recombination rates calculated by Nahar et al. (1997).
The letters `S' and `B' mark the ratios observed for the 23~Ori SLV components and the $\zeta$~Oph B components, respectively; the bar at right shows the range for each ratio observed in the local Galactic ISM.}
\label{fig:pred}
\end{figure}

Fig.~\ref{fig:pred} shows the ratios \mbox{X\,{\sc i}}/\mbox{Na\,{\sc i}} (for X = Mg, Ca, and Fe) that would be predicted for equilibrium between ionization (radiative plus collisional) and recombination (radiative plus dielectronic), as functions of temperature (as in Fig.~11 of Welty et al. 1999a, for example).
Those earlier calculations (which assumed `warm cloud' depletions) have been adjusted here for total gas phase elemental abundances characterized also by the patterns seen in the solar system (Lodders 2003) and in interstellar clouds exhibiting `cold cloud' depletions (e.g., Savage \& Sembach 1996; Welty et al. 1999b; see Table~\ref{tab:depl}).
The first six lines of Table~\ref{tab:ratios} list the predicted ratios for several representative combinations of depletion and temperature.
For Mg, Cr, and Fe -- for which the first ions \mbox{X\,{\sc ii}} should be dominant -- the predicted ratios are given by
\begin{equation}
\frac{\mbox{X\,{\sc i}}}{\mbox{Na\,{\sc i}}} = \frac{{\rm A}_{\rm X}~\delta_{\rm X}~\alpha_{\rm X}}
                                                  {{\rm A}_{\rm Na}~\delta_{\rm Na}~\alpha_{\rm Na}}
                                               \frac{(\alpha_{\rm Na}n_e + \Gamma_{\rm Na} + {\rm c}_{\rm Na}n_e)}
                                                    {(\alpha_{\rm X}n_e + \Gamma_{\rm X} + {\rm c}_{\rm X}n_e)},
\end{equation}
where A$_{\rm X}$ and $\delta_{\rm X}$ are the solar abundance and depletion of element X, respectively.
The calculation for calcium is slightly more complicated, as \mbox{Ca\,{\sc iii}} must also be considered.
The adopted photoionization rates ($\Gamma_{\rm X}$) are taken in most cases from P\'{e}quignot \& Aldrovandi (1986), assuming the WJ1 radiation field (de Boer et al. 1973); the collisional ionization rates (c$_{\rm X}$) are taken from Shull \& Van Steenberg (1982).
The total (radiative plus dielectronic) recombination rate coefficients ($\alpha_{\rm X}$) are calculated from the parameters given by Aldrovandi \& P\'{e}quignot (1973, 1974), Shull \& Van Steenberg (1982), and/or P\'{e}quignot \& Aldrovandi (1986).
The lower set of curves for \mbox{Fe\,{\sc i}} shows the ratios predicted using the photoionization and (total) recombination rates computed by Nahar, Bautista, \& Pradhan (1997).
For \mbox{Cr\,{\sc i}}, $\Gamma_{\rm Cr}$ = 8 $\times$ 10$^{-10}$ s$^{-1}$ was estimated by Meyer \& Roth (1990), and $\alpha_{\rm Cr}$ has been set to 6 $\times$ 10$^{-12}$ cm$^{3}$s$^{-1}$ (for $T$ = 100 K), similar to the values determined for radiative recombination to a number of other trace neutral species (P\'{e}quignot \& Aldrovandi 1986); the temperature dependence of $\alpha_{\rm Cr}$ is assumed to be the same as for $\alpha_{\rm Na}$ (i.e., with no significant contribution from dielectronic recombination for $T$ $<$ 12000 K; see comments below, however).
If the recombination is dominated instead by charge exchange with large molecules or small grains (Lepp et al. 1988; Weingartner \& Draine 2001; Liszt 2003; Welty et al. 2003), and if the resulting neutrals do not stick to the grains (sticking parameter $s$ = 0), then the predicted ratios change by at most 0.1 dex.

The predicted ratios for \mbox{Mg\,{\sc i}}, \mbox{Ca\,{\sc i}}, and \mbox{Fe\,{\sc i}} shown in Fig.~\ref{fig:pred} exhibit some common trends.
At temperatures below about 3000 K, the ratios reflect primarily photoionization and radiative recombination -- and are not very sensitive to the overall strength of the radiation field, the electron density $n_e$, or the temperature.
Because \mbox{Ca\,{\sc ii}} is often a trace species, however, the \mbox{Ca\,{\sc i}}/\mbox{Na\,{\sc i}} ratio does depend on $n_e$.
At somewhat higher temperatures, all the ratios increase due to dielectronic recombination (which does not significantly affect \mbox{Na\,{\sc i}} until much higher temperatures) -- starting at about 4000 K, 3000 K, and 5000 K for \mbox{Mg\,{\sc i}}, \mbox{Ca\,{\sc i}}, and \mbox{Fe\,{\sc i}}, respectively (Aldrovandi \& P\'{e}quignot 1973, 1974; Shull \& Van Steenberg 1982).
 
For the main cloud(s) toward $\zeta$ Oph, the observed ratios for \mbox{Mg\,{\sc i}}, \mbox{Ca\,{\sc i}}, and \mbox{Cr\,{\sc i}} are in good agreement with the predicted cold cloud values (consistent with the depletions derived from the corresponding dominant species and the temperature inferred from H$_2$ rotational excitation).
The ratio for \mbox{Fe\,{\sc i}} is lower than the predicted cold cloud value by more than a factor of 10, however [but is not as deficient relative to the value predicted using the Nahar et al. (1997) rates].
Meyer \& Roth (1990) remarked that the electron density estimated toward $\zeta$ Oph from the \mbox{Cr\,{\sc i}}/\mbox{Cr\,{\sc ii}} ratio was similar to the values obtained from other neutral/first ion ratios; Welty et al. (2003) noted that it is not uncommon for \mbox{Fe\,{\sc i}} to be somewhat weaker than expected, relative to other trace neutral species.
For the `strong low-velocity' clouds toward 23 Ori (where some trace neutral species are relatively strong and the depletions are intermediate between the representative warm and cold cloud values; Welty et al. 1999b) and for the main cloud(s) toward HD~72127A (with warm cloud depletions), the observed ratios for \mbox{Mg\,{\sc i}} and \mbox{Ca\,{\sc i}} are closer to the predicted warm cloud values for $T$ $\sim$ 3000 K (although the gas is likely much cooler than that in both cases), but the ratios for \mbox{Fe\,{\sc i}} again are low by factors of 10--25.
For the main cloud(s) toward HD~72127B, the observed ratios for both \mbox{Ca\,{\sc i}} and \mbox{Fe\,{\sc i}} are about a factor of 10 higher than those toward HD~72127A.
The ratio for \mbox{Ca\,{\sc i}} thus is closer to the value predicted for solar relative abundances than to that predicted for warm cloud abundances (for $T$ $\sim$ 3000 K), while the ratio for \mbox{Fe\,{\sc i}} is consistent with the predicted warm cloud value at that $T$ (i.e., low compared to \mbox{Ca\,{\sc i}}); the observed ratio for \mbox{Cr\,{\sc i}}, however, is {\it a factor of 10 higher than the predicted solar value}.

As calcium, chromium, and iron usually are much more severely depleted into dust than is sodium, an enhancement of the neutral species \mbox{Ca\,{\sc i}}, \mbox{Cr\,{\sc i}}, and \mbox{Fe\,{\sc i}}, relative to \mbox{Na\,{\sc i}}, could be due to much less severe depletions.
As noted above, dielectronic recombination can increase the abundances of \mbox{Mg\,{\sc i}}, \mbox{Ca\,{\sc i}}, and \mbox{Fe\,{\sc i}} when the temperature exceeds about 4000 K, 3000 K, and 5000 K, respectively.
The cloud at $+$65 km~s$^{-1}$ toward SN1987A is relatively cold [as the \mbox{Na\,{\sc i}} D lines at that velocity exhibit resolved hyperfine structure, with $b$ $\sim$ 0.3 km~s$^{-1}$ (Pettini \& Gillingham 1988; Welty \& Crowther, in preparation)], but the ratios of various dominant species suggest that there is essentially no depletion there (Welty et al. 1999a).
The observed \mbox{Ca\,{\sc i}}/\mbox{Na\,{\sc i}} ratio is within a factor of 2 of the predicted value for solar relative abundances at 100 K (i.e., a cold, dust-free cloud), though it is also consistent with the values predicted for somewhat warmer gas.
For the component at $+$3 km~s$^{-1}$ toward $\epsilon$~Ori, the $b$-values for \mbox{Na\,{\sc i}}, \mbox{Ca\,{\sc i}}, and \mbox{Ca\,{\sc ii}} -- all 0.36-0.38 km~s$^{-1}$ -- indicate that that cloud also is fairly cold (Welty et al. 1994, 1996, 2003).
The slightly elevated \mbox{Mg\,{\sc i}}/\mbox{Na\,{\sc i}} ratio thus may reflect less severe depletion of magnesium there (and not dielectronic recombination).
The still higher values of \mbox{Ca\,{\sc i}}/\mbox{Na\,{\sc i}} and/or \mbox{Fe\,{\sc i}}/\mbox{Na\,{\sc i}} seen in several other cases, however, would seem to imply some combination of mild (or negligible) depletions and dielectronic recombination in warmer gas (e.g., Gnaci\'{n}ski \& Krogulec 2008)\footnotemark -- though the \mbox{Ca\,{\sc i}}/\mbox{Ca\,{\sc ii}} ratios generally are not unusually high.
\footnotetext{Gnaci\'{n}ski \& Krogulec (2008) assume that there is no depletion in the CaFe clouds -- taking non-detections of CH to imply a lack of both H$_2$ and dust.  
We note, however, that there are many sight lines where CH is not detected (to similar limits), but which none the less have $E(B-V)$ greater than zero, detections of H$_2$ (at column densities below the regime of self-shielding), and evidence from dominant species for at least moderate depletions.
The high temperatures ($\ga$ 8000 K) derived for the CaFe clouds presumably reflect the ability of dielectronic recombination to produce the observed abundances of \mbox{Ca\,{\sc i}} and \mbox{Fe\,{\sc i}} in such warm gas.
Allowance for depletions and for colder temperatures, however, could reduce the substantial discrepancies between their models and the \mbox{Na\,{\sc i}} and \mbox{K\,{\sc i}} observed for the `non-CaFe' clouds.}

The strong observed absorption from \mbox{Ca\,{\sc i}}, \mbox{Cr\,{\sc i}}, and \mbox{Fe\,{\sc i}} toward HD~72127B -- relative to \mbox{Na\,{\sc i}} and to the much weaker absorption from those species toward HD~72127A -- also may suggest some combination of mild depletions and higher temperature.
The similarities in the strengths of the 5780 \AA\ DIB, the $E(B-V)$ colour excesses, and the equivalent widths of the weak \mbox{Mg\,{\sc ii}} $\lambda$1239,1240 lines toward HD~72127A and HD~72127B suggest that the total hydrogen column densities are similar for the two sight lines.
The relatively small differences in the column densities of \mbox{Na\,{\sc i}} and \mbox{Ti\,{\sc ii}} in the two sight lines (both overall and in the main components near +12 km~s$^{-1}$) and in the \mbox{Ca\,{\sc i}}/\mbox{Ca\,{\sc ii}} ratios (in the main components) then suggest that differences in ionization and depletions could account for only part of the enhancement of \mbox{Ca\,{\sc i}}, \mbox{Cr\,{\sc i}}, and \mbox{Fe\,{\sc i}}.
The higher values for \mbox{Ca\,{\sc i}}/\mbox{Na\,{\sc i}} and \mbox{Fe\,{\sc i}}/\mbox{Na\,{\sc i}} toward HD~72127B could be explained, however, if the \mbox{Ca\,{\sc i}} and \mbox{Fe\,{\sc i}} were enhanced via dielectronic recombination, in gas with warm cloud depletions and $T$ $\ga$ 5000 K. 
We conjecture that dielectronic recombination to \mbox{Cr\,{\sc i}} might then account for the high \mbox{Cr\,{\sc i}}/\mbox{Na\,{\sc i}} ratio toward HD~72127B -- and we would predict some enhancement of \mbox{Mg\,{\sc i}} (which is not unusually strong toward HD~72127A), but no enhancement of \mbox{Si\,{\sc i}} (for which dielectronic recombination becomes important at somewhat higher temperatures)\footnotemark. 
If dielectronic recombination is responsible for the strong absorption from those trace neutral species toward HD~72127B, then there would be a significant difference in temperature for the main components toward HD~72127A and HD~72127B -- over a distance of less than 2200 AU.
We note, however, that the \mbox{Ca\,{\sc i}}/\mbox{Ca\,{\sc ii}} ratio often seems to imply a higher $n_e$ than other such trace/dominant ratios -- even when the \mbox{Ca\,{\sc i}} $b$-values place strong constraints on $T$ (Welty et al. 2003) -- so that high \mbox{Ca\,{\sc i}} abundances are not necessarily due to dielectronic recombination in warm gas.
\footnotetext{After the initial submission of this paper, we received preliminary, partial calculations of the total recombination rate coefficients to \mbox{Cr\,{\sc i}} (S. Nahar, private communication), which both confirm our assumptions regarding the recombination at $T$ $\la$ 1000 K and suggest that dielectronic recombination may indeed be significant below 10$^4$ K for \mbox{Cr\,{\sc i}}.}

\subsection{Location of gas}
\label{sec-loc}

The location of the main absorption components at $v$ $\sim$ $+$10 to $+$13 km~s$^{-1}$ ($v_{\rm LSR}$ $\sim$ $-$3 to 0 km~s$^{-1}$) toward HD~72127AB -- with depletions apparently similar to those found in typical warm, diffuse disc clouds -- is not known.
The outlying components, with higher LSR velocities and much milder depletions, are very likely associated with the Vela SNR, at a distance of about 250$\pm$30 pc (Cha et al. 1999).
Some absorption at low LSR velocities, however, is seen for stars closer than 200 pc -- and appears to be located at distances less than about 130 pc (Cha et al. 2000).
The foreground absorption is `patchy and inhomogeneous' (Cha et al. 2000), and the highest column density of \mbox{Na\,{\sc i}} seen toward any star foreground to the SNR is about 7.2$\times$10$^{11}$ cm$^{-2}$ (the $v$ = 12.3 km~s$^{-1}$ component toward HD~72232, at less than 115 pc) -- a factor of 3--5 lower than the values obtained for the main components toward HD~72127AB.
The relatively weak \mbox{Ca\,{\sc ii}} absorption toward HD~72232 (15 m\AA; Cha \& Sembach 2000) implies a \mbox{Ca\,{\sc ii}} column density of about 2$\times$10$^{11}$ cm$^{-2}$; the corresponding \mbox{Ca\,{\sc ii}}/\mbox{Na\,{\sc i}} ratio, about 0.3, is similar to the value found for the main component toward HD~72127A (Table~\ref{tab:comps}).
The much higher column densities and the continuing variations in both column density and velocity suggest that the bulk of the material in the main components toward HD~72127AB is associated with the SNR, but there is very likely some contribution from foreground material (at the same velocity) as well.

%%%%%%%%%%%%%%%%%%%%%%%%%%%%%%%%%%%%%%%% SUMMARY %%%%%%%%%%%%%%%%%%%%%%%%%%%%%

\section{Summary / Conclusions}
\label{sec-summ}

New, moderately high resolution (FWHM = 4.5--4.9 km~s$^{-1}$) optical spectra of \mbox{Na\,{\sc i}} and \mbox{Ca\,{\sc ii}} absorption toward HD~72127AB have provided additional evidence for both spatial and temporal variations in the complex interstellar absorption along the two sight lines.
If the spatial differences are associated with material in the Vela SNR, they occur on scales of about 1100 AU.
Fits to the absorption-line profiles seen in high-resolution (FWHM $\sim$ 3.5 km~s$^{-1}$) UV spectra of HD~72127A obtained with the {\it HST} GHRS have yielded abundances for a number of species (e.g., \mbox{Si\,{\sc ii}}, \mbox{Fe\,{\sc ii}}, \mbox{Ni\,{\sc ii}}, and \mbox{Zn\,{\sc ii}}) in the various velocity components discernible in the spectra.
The main components at low LSR velocities have depletions similar to those found in warm, diffuse disc clouds; the generally weaker components at higher velocities have much milder depletions, more similar to those found in halo clouds.
Similarities in $E(B-V)$ and in the equivalent widths of the 5780 \AA\ diffuse interstellar band and the UV lines of \mbox{Mg\,{\sc ii}} and \mbox{S\,{\sc ii}} suggest that the total hydrogen column densities in the two sight lines are similar, with log[$N$(H) (cm$^{-2}$)] $\sim$ 20.40.
Similarities in the column densities of \mbox{Ti\,{\sc ii}} and \mbox{Na\,{\sc i}} suggest that the depletions and ionization in the main components also are comparable (within factors of about 2).
Several other trace neutral species -- \mbox{Ca\,{\sc i}}, \mbox{Cr\,{\sc i}}, and \mbox{Fe\,{\sc i}} -- are much stronger toward HD~72127B, however.
In particular, the column density of \mbox{Cr\,{\sc i}} is about 30 times the value found for the main components toward $\zeta$~Oph (the only other sight line in which \mbox{Cr\,{\sc i}} has been detected) -- even though $N$(H) is a factor of about 6 lower toward HD~72127B.
While an earlier high-resolution spectrum of \mbox{Ca\,{\sc i}} toward HD~72127A indicates that the main component in that sight line has $T$ $\la$ 900 K, the strong absorption from \mbox{Ca\,{\sc i}} and \mbox{Fe\,{\sc i}} toward HD~72127B suggests that the main component there is much warmer -- $T$ $\ga$ 5000 K -- with dielectronic recombination largely responsible for the enhanced abundances of those neutral species.
We conjecture that dielectronic recombination may also be responsible for the enhanced \mbox{Cr\,{\sc i}} absorption toward HD~72127B, and predict that strong absorption will be found for \mbox{Mg\,{\sc i}} as well.

It would be very useful to obtain high-resolution UV spectra of HD~72127B:  
\mbox{Si\,{\sc ii}}, \mbox{Cr\,{\sc ii}}, \mbox{Fe\,{\sc ii}}, and \mbox{Zn\,{\sc ii}} [to determine $N$(H) and depletions and to compare with HD~72127A];
\mbox{Mg\,{\sc i}} and \mbox{Si\,{\sc i}} (to obtain constraints on the effect of dielectronic recombination); and
\mbox{C\,{\sc i}} (also toward HD~72127A; to estimate thermal pressures and densities in the various components).
Higher resolution optical spectra of the trace neutral species would provide more stringent constraints on the temperature in the main component.
Determination of the recombination behaviour of \mbox{Cr\,{\sc ii}}, as a function of temperature, would aid in understanding the high abundance of \mbox{Cr\,{\sc i}} and the role of dielectronic recombination.
Continued monitoring of the interstellar absorption toward these two stars could provide insights into the structure and properties of interface regions between warm and cool gas.

%%%%%%%%%%%%%%%%%%%%%%%%%%%%%%% ACKNOWLEDGMENTS %%%%%%%%%%%%%%%%%%%%%%%%%%%%%%

\section*{Acknowledgments}

We are grateful to M. Rejkuba and M.-R. Cioni (ESO/Paranal) for their assistance with the UVES observations, to J. Thorburn for performing initial processing of the UVES data, and to S. Nahar for communicating preliminary results on the recombination to \mbox{Cr\,{\sc i}}.
T. S. acknowledges support from REU grant NSF-0353854 to the University of Chicago.
This work has been supported by NASA Long-Term Space Astrophysics grants NAGW-4445 and NAG5-11413 to the University of Chicago.
The GHRS data for HD~72127A were originally obtained under NASA grant GO-2251.01-87A to the University of Chicago.

%%%%%%%%%%%%%%%%%%%%%%%%%%%%%%%%%%%%%% APPENDICES %%%%%%%%%%%%%%%%%%%%%%%%%%%%

\end{document}